\documentclass{emulateapj}
\usepackage{epsf}
\usepackage{amsmath,graphicx}
\usepackage[z]{tmsfnts}

\newcommand{\nbody}{{\tt DMO}}
\newcommand{\ad}{{\tt DMG\_NR}}
\newcommand{\csf}{{\tt DMG\_SF}}

\newcommand{\cloudy}{{\tt CLOUDY}~}

\newcommand{\mpch}{h\, {\mathrm{Mpc}^{-1}}}
\newcommand{\hmpc}{h^{-1}\, {\mathrm{Mpc}}}
\newcommand{\hkpc}{h^{-1}\, {\mathrm{kpc}}}
\newcommand{\hmsun}{\ h^{-1}\, {\mathrm{M}}_{\odot}}

\newcommand{\Rvir}{R_{\mathrm{vir}}}

\newcommand{\dvir}{\Delta_{\mathrm{vir}}}

\newcommand{\cburk}{c_{\mathrm{B}}}
\newcommand{\Msun}{M_{\odot}}
\newcommand{\dd}{\mathrm{d}}

\newcommand{\LCDM}{\Lambda{\mathrm{CDM}}}

\newcommand{\Omegam}{\Omega_{\mathrm{M}}}
\newcommand{\Omegab}{\Omega_{\mathrm{B}}}
\newcommand{\Omegal}{\Omega_{\mathrm{\Lambda}}}
\newcommand{\sige}{\sigma_{\mathrm{8}}}

\newcommand{\Si}{\mathrm{Si}}
\newcommand{\Ci}{\mathrm{Ci}}

\begin{document}
\bibliographystyle{apj}

\slugcomment{{\em The Astrophysical Journal, submitted}} 
\shortauthors{RUDD, ZENTNER, \& KRAVTSOV}
\shorttitle{BARYONS AND THE POWER SPECTRUM}

\title{Effects of Baryons and Dissipation on the Matter Power Spectrum}

\author{
Douglas H. Rudd\altaffilmark{1}, 
Andrew R. Zentner\altaffilmark{1,2,3},
\& Andrey V. Kravtsov\altaffilmark{1,2}
}
\altaffiltext{1}{Department of Astronomy \& Astrophysics \& Kavli Institute for Cosmological 
Physics, The University of Chicago, Chicago, IL 60637 USA}
\altaffiltext{2}{Enrico Fermi Institute, The University of Chicago, Chicago, IL 60637 USA }
\altaffiltext{3}{National Science Foundation Astronomy and Astrophysics Postdoctoral Fellow}

\begin{abstract}

We study the importance of baryonic physics on predictions of the matter 
power spectrum as it is relevant for forthcoming weak lensing surveys.  
We quantify the impact of baryonic physics using a set of three cosmological 
numerical simulations.  Each simulation has the same initial density field, 
but models a different set of physical processes.  The first 
simulation evolves the density field using gravity alone, the second includes 
non-radiative gasdynamics, and the third includes radiative heating and cooling 
of baryons, star formation, and supernova feedback.  We find that baryonic processes 
alter predictions for the matter power spectrum significantly relative to  
models that include only gravitational interactions.  Our results imply that 
future weak lensing experiments such as LSST and SNAP will be very sensitive to 
the poorly-understood physics governing the nonlinear evolution of the baryonic 
component of the universe.  The net effect is significantly larger in the case of 
the model with cooling and star formation, in which case our results imply that 
contemporary surveys such as the CFHT Wide survey may also be sensitive to 
baryonic processes.  In particular, this effect could be important 
for forecasts of the constraining power of future surveys if information from
scales $\ell \gtrsim 1000$ is included in the analysis.
We find that deviations are caused primarily by the 
rearrangement of matter within individual dark matter halos relative to 
the gravity-only case, rather than a large-scale rearrangement of matter.  
Consequently, we propose a simple model, based on the phenomenological 
halo model of dark matter clustering, 
for baryonic effects that can be used to aid 
in the interpretation of forthcoming weak lensing data.

\end{abstract}


\section{Introduction}
\label{sec:intro}

Contemporary determinations of the cosmic energy budget using a variety of 
cosmological probes all indicate that the majority of the energy 
in the universe ($\sim 70\%$) is in the form of {\em dark energy} 
with negative pressure that drives an accelerated cosmic expansion 
\citep[{\it e.g.,}~][]
{riess_etal98,perlmutter_etal99,tegmark_etal04,riess_etal04,eisenstein_etal05,spergel_etal06,tegmark_etal06,astier_etal06,wood-vasey_etal07}.
The unknown nature and properties of the dark energy is widely recognized as one of the 
most important and fundamental problems in cosmology, if not all of physics.

Future large surveys, such as the Dark Energy Survey\footnote{{\tt
http://www.darkenergysurvey.org/}} (DES), the SuperNova/Acceleration
Probe\footnote{{\tt http://snap.lbl.gov}}
\citep[SNAP,][]{aldering_etal04}, and the Large Synoptic Survey
Telescope\footnote{{\tt http://www.lsst.org}} (LSST), are expected to
measure the matter density fluctuation statistics from the linear to
the nonlinear regime with unprecedented precision
\citep[{\it e.g.,}~][]{aldering05,tyson05}.  The statistical power of such
precision measurements should allow for stringent constraints of the
properties of dark energy
\citep[{\it e.g.,}~][]{hu_tegmark99,hu99,huterer02,heavens03,refregier03,refregier_etal04,song_knox04,takada_jain04,takada_white04,dodelson_zhang05,albrecht_etal06,zhan06}.
However, realizing this goal requires theoretical predictions for the
power spectrum in the linear and nonlinear regimes and as a function of
cosmology with an accuracy of order a percent 
\citep[{\it e.g.,}~][]{huterer_takada05,huterer_etal06}.

The theoretical aspect of this program is daunting.  To realize fully 
these goals, the power spectrum must be calibrated on scales where
nonlinear effects are important.  Consequently, the calibration
program relies on numerical simulations of cosmological structure
formation \citep[{\it e.g.,}~][]{annis_etal05}.  However, it has not been yet
demonstrated that modern cosmological simulations can achieve the
required accuracy in the nonlinear regime. On the contrary, recent
systematic studies show differences of order $\sim 10\%$ 
between different simulation codes at the relevant range of scales 
\citep[$k\approx 1-10\rm\,Mpc^{-1}$;][]{heitmann_etal05}. 
For example, the limited numerical resolution of simulations, and therefore the
limited ability of simulations to resolve the inner structures of dark
matter halos and the properties of halo substructure, can give rise to
non-negligible systematic effects on predicted power spectra
\citep[{\it e.g.,}~][]{hagan_etal05}.

So far, systematic studies of the nonlinear power spectrum, 
including the calibration of popular analytic, phenomenological 
models, such as the halo model \citep[see][for a review]{cooray_sheth02}, 
have utilized dissipationless 
$N$-body simulations of gravitational clustering 
\citep[{\it e.g.,}~][]{smith_etal03}.  
However, in the observed Universe approximately $\sim 15\%$ 
of matter is in the form of baryons \citep{spergel_etal06}.  
Although baryons are sub-dominant and trace the distribution 
of dark matter at the onset of structure growth, 
the final distribution of baryons in halos differs 
from that of dark matter significantly because 
they differ in their subsequent dynamical evolution.

This differentiation has been observed routinely in 
cosmological simulations of the formation of galaxy clusters.  
In general, these studies show that baryons have a 
more extended distribution than that of dark matter 
\citep[{\it e.g.,}~][]{frenk_etal99} and 
this result is borne out by observations 
\citep[{\it e.g.,}~][]{david_etal95,vikhlinin_etal06}. 
Recent simulations \citep[{\it e.g.,}~][]{rasia_etal04,lin_etal06} have shown
that the transfer of energy from dark matter to baryons
can modify the concentration of the dark matter radial 
distribution by $\approx 10\%$, even in simulations
that do not include baryonic dissipation.  Baryons may 
also be redistributed from high- to low-density regions 
via energetic AGN feedback, giving rise to another, potentially 
non-negligible, effect on the matter power spectrum \citep{levine_gnedin06}.

Furthermore, baryons dissipate energy through radiative processes,
leading to the condensation of baryons in the central regions of halos
where gas densities sufficient for star formation can be achieved.
The condensing baryons pull dark matter along with them, leading to an
increase in the dark matter density
\citep[{\it e.g.,}~][]{zeldovich_etal80,blumenthal_etal86,gnedin_etal04,sellwood_mcgaugh05}.
As a result, in any treatment of structure growth that
self-consistently includes the condensation of baryons and the
formation of galaxies, not only will some fraction of the total mass
be redistributed into galaxies with comparably small spatial extent,
but the dark matter halos themselves will have more compact, concentrated 
internal mass distributions.

Previous analytic estimates of the effects of baryon dissipation on the matter
power spectrum \citep[{\it e.g.,}~][]{white04,zhan_knox04} suggest that the
effect should be confined to small scales.  However, these preliminary
studies are based on approximate, phenomenological models and may not
have taken all of the relevant effects into account.  For instance,
\citet{zhan_knox04} model the relative influence of the hot baryons
of the intracluster medium only, and neglect the influence of a cold
component.  

A recent study by \citet{jing_etal06} uses numerical simulations
to model self-consistently the interaction between baryons and dark matter
and shows that dissipational physics 
can affect the power spectrum on scales $k \sim 1-10\mpch$ by 
$\gtrsim 5-10\%$. Our results provide qualitative confirmation of
this result.  Moreover, as we show below, in addition to the well-known
increase of dark matter density in halo centers in response to baryon
condensation, an effect included in the calculation of
\citet{white04}, simulations indicate associated changes in the
distribution of matter at larger radii. This effect leads to enhanced concentration of the
overall matter distribution in halos and affects the power spectrum on 
scales larger than previously thought.

In the present paper, we use high-resolution, cosmological simulations
of structure formation to study the effect of baryons and dissipation
on the power spectrum of matter fluctuations at scales $k \sim 0.1-10
\mpch$.  In particular, we perform a series of three
simulations each of which begins with the same set of initial
conditions but includes different matter components and physical
processes.  The first is a collisionless $N$-body simulation that 
models only a dark matter component and therefore includes only gravitational 
interactions.  In the second simulation, we include a 
baryonic component evolving according to an Eulerian 
hydrodynamics method.  In this second simulation, the 
baryonic component is not permitted to cool radiatively.  In the
third simulation, we include radiative cooling for the baryonic
component and prescriptions for star formation and feedback processes.
We use the results of these simulations to study the relative impact
of baryonic physics on the matter power spectrum and use the analytic 
halo model to show how these effects of baryons on
the power spectrum can be understood
in terms of differences in the density distributions of baryons and
dark matter in simulations with baryonic physics.  We conclude with a
brief discussion of the implications of our results for upcoming
precision weak lensing measurements of the matter power spectrum. For
the purpose of inferring the properties of dark energy from weak lensing
measurements, we suggest that it may be possible to encapsulate the
effects of galaxy formation in a small number of parameters.

The paper is organized as follows. In \S~\ref{sec:methods}
we describe the set of cosmological simulations used in this study and in 
\S~\ref{sec:results}, we describe the results of 
our analysis.  These results include both three-dimensional 
and convergence power spectra as well as the structural properties 
of the dark matter halos and corresponding 
baryonic components in the simulations.  
We describe the halo model for matter clustering in 
\S~\ref{section:halomodel} and use it
to aid in the interpretation of the simulation results 
of \S~\ref{sec:results}.  In \S~\ref{section:halomodel}, 
we also suggest a method, based on the halo model, 
that can be used to model the baryonic 
physics present in our simulations.  
We summarize our conclusions and discuss their 
implications in \S~\ref{s:discussion}.

\section{Simulations and Analysis Methods}
\label{sec:methods}

We investigate the influence of baryons on the matter power spectrum
using cosmological numerical simulations.  We simulate the formation
of structure in a cubic volume $60 \hmpc$ on a side in the
``concordance'' $\LCDM$ cosmology ($\Omegam = 0.3$, $\Omegal = 0.7$,
$\Omegab h^2= 0.021$, $h = 0.7$, $\sige = 0.9$).  

We use a set of three simulations, each of which begins
with the same initial density field but includes different physical
processes.  The first simulation, ``$\nbody$,'' is purely
dissipationless and includes a collisionless dark matter component only.
The second simulation, ``$\ad$,'' follows both dark matter and
baryonic components but does not include radiative cooling for the
latter.  The gas in this run is thus modeled in the non-radiative (or
``adiabatic'') regime.  The third simulation, ``$\csf$,'' treats the
baryonic component including radiative cooling and heating, star
formation, and feedback from supernovae. The inclusion of these physical
processes allows for the formation of galaxies in the $\csf$
simulation as cooled gas forms a condensed component, a fraction of
which is converted into stars. A basic description of these simulations
is provided in Table~\ref{tab:sim}.

\begin{deluxetable}{lcccc}
\tablecolumns{5}
\tablecaption{List of Simulations\label{tab:sim}}
\tablehead{\multicolumn{1}{c}{Simulation}&
\multicolumn{1}{c}{Dark Matter} &
\multicolumn{1}{c}{Baryons} &
\multicolumn{1}{c}{Cooling/SF} &
\multicolumn{1}{c}{Resolution} }
\startdata
$\nbody$ & {yes} & {no}  & {no}  & $0.9 \hkpc$ \\
$\ad$    & {yes} & {yes} & {no}  & $1.8 \hkpc$ \\
$\csf$   & {yes} & {yes} & {yes} & $3.6 \hkpc$\\
\enddata
\end{deluxetable}

\begin{figure}
\plotone{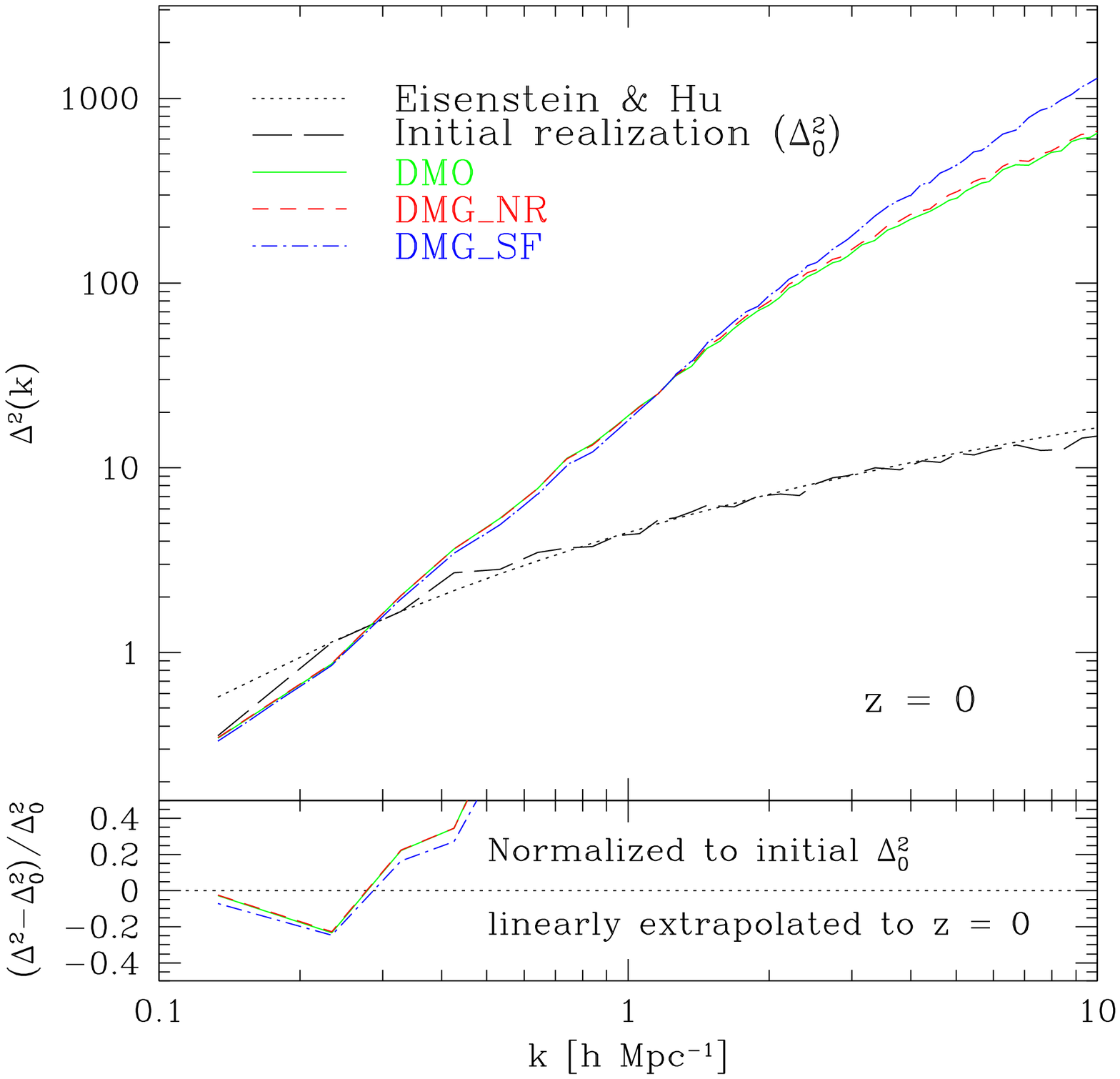}
\caption{{\it Upper Panel}: The simulated power spectra for the $N$-body ($\nbody$),
non-radiative ($\ad$) and cooling ($\csf$) simulations at $z = 0$, as well
as the linear power spectrum obtained from the transfer function of 
\citet{eisenstein_hu99} and the actual initial density fluctuation spectrum
used in the simulations.  Note that the deficit of power relative to the 
\citet{eisenstein_hu99} spectrum is partially due to our initial realization,
and partially due to nonlinear evolution.
{\it Lower Panel}: The simulated power spectra now plotted relative to
the initial density fluctuation spectrum.  Note that this panel shows the 
power spectrum normalized to the initial power spectrum realization used 
in our simulations and is not a residual compared to the \citet{eisenstein_hu99}
spectrum. 
\label{fig:pk}}
\end{figure}

As the simulation volume is fairly small, cosmic variance and finite
volume effects are significant at scales corresponding to 
$k \sim 0.1-10 \mpch$.  This fundamental limitation 
prevents us from presenting {\it absolute} estimates 
of the power spectrum at precisions of order a percent.
Additionally, we detect the effect of limited volume on the largest
scales, $k \sim 0.25 \hmpc$, which is close to the scale at which
density fluctuations have become nonlinear by $z = 0$ (see
Figure~\ref{fig:pk}).  The transition to nonlinearity occurs between
the fundamental mode and this scale, so the finite box size may also
affect the growth of the fundamental mode.  However, these effects
should be similar in all three simulations, and we are
primarily interested in the {\em relative} effects of baryonic
processes, which should be discernible through a comparison of the
$\nbody$, $\ad$, and $\csf$ simulations.

We perform the dissipationless simulation ($\nbody$) using the Adaptive
Refinement Tree (ART) $N$-body code
\citep[]{kravtsov_etal97,kravtsov99}.  The ART code employs adaptive
refinement in space and time in order to achieve the large dynamic
range necessary to model the detailed structure of dark matter halos.
In the $\ad$ and $\csf$ simulations, we follow the evolution of the 
gaseous baryonic component using an Eulerian hydrodynamics 
solver on the same adaptive mesh of the $N$-body ART code, 
as described by \citet{kravtsov_etal02}. However, we perform 
the two simulations that include baryons utilizing the new, 
distributed-memory version of the 
$N$-body+gasdynamics ART code (Rudd \& Kravtsov, in preparation).

The simulations with baryons treat both the collisionless
gravitational dynamics of the dark matter and stars as well as
hydrodynamical evolution of the gas component.  Hydrodynamic fluxes
are computed at the boundary of mesh cells with a second order
Godunov-type Riemann solver, which is particularly efficient at
resolving shocks (typically within $\approx$1-2 grid cells).  The
simulation with radiative cooling treats several additional physical
processes, including star formation, metal enrichment and thermal
feedback due to supernovae Type II and Type Ia, self-consistent
advection of metals, metallicity-dependent radiative cooling, and UV
heating due to a cosmological ionizing background
\citep{haardt_madau96}.  The cooling and heating rates take into
account Compton heating and cooling of the plasma, UV heating, and
atomic and molecular cooling.  The cooling rates are tabulated
for the temperature range $10^2<T<10^9$~K on a grid of metallicities
and UV intensities using the \cloudy code
\citep[ver. 96b4;][]{ferland_etal98}.  The \cloudy cooling and heating
rates take into account the metallicity of the gas, which is
calculated self-consistently during the simulation, so that the local
cooling rates depend on the local gas metallicities.

Star formation in the $\csf$ simulation is implemented according to the 
observationally-motivated recipe \citep[{\it e.g.,}~][]{kennicutt98}:
$\dot{\rho}_{\ast}\propto\rho_{\rm gas}^{1.5}/t_{\ast}$, with
$t_{\ast}=4\times 10^9$~yrs.  
The gas in mesh cells is converted into collisionless
stellar particles stochastically on timescales 
$\tau_{\mathrm{SF}} = 10^{8}$~yrs, where the mass of the particle 
created from the gas is given by 
$m_{\ast} = \dot{\rho}_{\ast} \tau_{\mathrm{SF}} V_{\mathrm{cell}}$, 
where $V_{\mathrm{cell}}$ is the volume of the cell.  
This prescription effectively averages 
the star formation rate in time, and is used
to prevent large numbers of low-mass stellar 
particles from forming in high-resolution cells that 
contain little mass owing to their small volumes.  
The code also accounts for stellar feedback on the surrounding gas, 
including the injection of energy and heavy elements (metals) 
via stellar winds, supernovae, and secular mass loss.

All three simulations follow the evolution of $256^3$ dark matter
particles (implying a particle mass of $m_p=1.07\times
10^9h^{-1}\,\Msun$ in $\nbody$, and $m_p=9.17\times 10^8h^{-1}\,\Msun$
in both $\ad$ and $\csf$).  Simulation $\nbody$ uses a $256^3$ uniform
root grid and up to of eight refinement levels,
giving a maximum comoving spatial resolution of $\approx 0.9 \hkpc$.
The simulations $\ad$ and $\csf$ each use a $128^3$ uniform root grid
and allow for eight and seven levels of refinement, resulting in 
minimum comoving cell sizes of of $1.8\hkpc$ and $3.6\hkpc$ on a side,
respectively.  The simulations with baryons use somewhat coarser
resolution compared to the $\nbody$ run to compensate for the
additional computational cost of these simulations, however the
resolution of all three simulations is small compared to the scales
we discuss.

We identify halos in the simulations using a variant of the Bound
Density Maxima algorithm (BDM, \citealt{klypin_etal99a}), as described
in \citet{kravtsov_etal04b}.  We use the publicly-available {\tt
smooth}
code\footnote{\tt{http://www-hpcc.astro.washington.edu/tools/smooth.html}}
to compute the local density at the position at each particle smoothed
with a 24-particle smoothed particle hydrodynamics (SPH) kernel.  We
then identify halo centers with peaks in the smoothed dark matter
density field.  All particles within a search radius of $r_{\mathrm{f}} = 50
\hkpc$ of a peak are removed from further consideration as potential
halo centers.  The value of the parameter $r_{\mathrm{f}}$ is set according to
the sizes of the smallest objects we aim to identify in the
simulations.  After identifying halos according to the BDM algorithm,
we calculate masses and other halo properties using all the matter
components present in the simulation (i.e., dark matter, gas, stars)
within a virial radius, $\Rvir$.  The virial radius is defined as the
radius of the sphere, centered on the highest-density particle in the
halo, within which the mean density is a contrast $\dvir(z)$ with
respect to the mean matter density of the Universe.  The value of
$\dvir(z)$ is set according to the spherical tophat collapse model and
can be computed quickly and accurately using the fitting formula of
\citet{bryan_norman98}.  In the concordance $\LCDM$ cosmology used in
this work, $\dvir(z=0) \approx 337$ and $\dvir \to 178$ for $z \gtrsim
1$.  In what follows, we consider only distinct halos, or halos that
do not lie within the virial radius of another, more massive halo.

We measure the power spectra in the simulation volumes using the
method of \citet[][see also
\citealt{kravtsov_klypin99}]{jenkins_etal98}.  The power spectrum is
computed in a series of wavenumber ranges $[k_{\rm i},k_{\rm i+1}]$ 
as follows.  We first divide the computational volume into $i^3$ cubic subvolumes.  
We superpose the particle distributions in each subvolume and assign
particle densities to points on a $512^3$ grid using this composite
particle distribution according to the cloud-in-cell interpolation
scheme \citep[]{hockney_eastwood81}.  For the gaseous component, we
assign densities directly from the adaptive refinement mesh.  The fast
Fourier transform of the density field then yields an estimate of the
power in modes that are periodic with period $L_{\rm box}/i$, where
$L_{\rm box} = 60 \hmpc$ is the comoving box size. 
The wavenumbers at the boundaries of each interval are set to $k_{\rm i} =
512 i (2\pi/L_{\rm box})/18$ in order to minimize the difference between
adjacent segments.  We use $i=2^m$ with $m = 0 \ldots 4$.  We make no
corrections for particle shot noise, which is negligible on all of the
scales that we consider.

\section{Simulation Results}
\label{sec:results}

\subsection{Power Spectra}
\label{sub:spectra}

\subsubsection{Three-Dimensional Matter Power Spectra}
\label{subsub:power3d}


\begin{figure}
\plotone{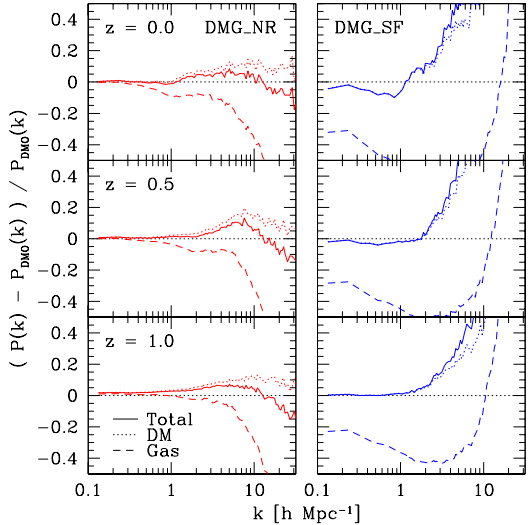}
\caption{
The simulated power spectra for the 
non-radiative ($\ad$) and cooling+star formation ($\csf$) simulations are
plotted in left and right columns respectively.  In each
case we plot the fractional difference of the power spectra
from that of the
dissipationless ($\nbody$) simulation.  We present results 
for a range of redshifts $z=0$, $0.5$, and $1.0$, as labeled in
the left panel of each row.  In each panel we show results for all matter
({\it{solid line}}) as well as for the dark matter 
({\it{dotted line}}) and gas ({\it{dashed line}}) components alone.  Stellar
mass in the ($\csf$) simulation is highly biased relative to
the total matter distribution and is omitted for clarity.
}
\label{fig:dpk}
\end{figure}

Figure~\ref{fig:dpk} shows the difference between 
the simulated power spectra in the radiative ($\csf$) and 
non-radiative ($\ad$) hydrodynamical simulations 
relative to the spectrum measured in the $N$-body 
simulation ($\nbody$). The spectra are plotted from 
the fundamental mode at $k=2\pi/L_{\rm box} \approx 0.1 \mpch$
to $k\approx 30 \mpch$ below which the power should 
not be affected directly by numerical resolution.

First, consider the power spectra from simulation $\ad$ with
non-radiative gas physics.  Relative to the $N$-body simulation
$\nbody$, the total matter power spectrum in the $\ad$ simulation 
is enhanced by of order $\sim 5-10 \%$ at
wavenumbers $1 \lesssim k/(\mpch) \lesssim 10$ at all redshifts.
Decomposing this change into the contributions from the gas and dark
matter components, we see that the dark matter power is enhanced by
$\sim 10-15\%$ on these scales.  Conversely, the gas component
exhibits diminished power relative to the $N$-body simulation on all
scales $k \gtrsim 0.7 \mpch$.  In the range of wavenumbers $1 \lesssim
k/(\mpch) \lesssim 6$, the power in the gas component is reduced by
roughly $\sim 10\%$ at all redshifts, but the power drops
precipitously for $k \gtrsim 6 \mpch$.  The net result is that the
total matter power spectrum in the non-radiative gas simulation is
enhanced on scales $1 \lesssim k/(\mpch) \lesssim 10$, but drops
rapidly for larger wavenumbers.  These results are in qualitative
agreement with the results of the non-radiative simulation presented
by \citet{jing_etal06}, but there are differences in detail.  In
particular, \citet{jing_etal06} report that the gas power spectrum is
reduced by $\sim 20\%$ or more on scales $1 \lesssim k/(\mpch) \lesssim 6$ 
at $z=0$.  Additionally, \citet{jing_etal06} report 
a slightly larger enhancement in the dark matter
power on these scales.  However, the net result is a total matter power
spectrum that differs from their $N$-body results by less than $\sim
5\%$ for $k \lesssim 6 \mpch$.  While these two results are similar,
it is important to note that they are achieved 
via different combinations of suppressed gas power and enhanced 
dark matter power.

Relative to the non-radiative simulation, the effect on the power
spectrum in the simulation with gas cooling and star formation
($\csf$) is significantly more dramatic.  In this case, the gas power
spectrum is greatly reduced on large scales ($k \lesssim 10 \mpch$)
relative to the $N$-body matter power spectrum.  This large-scale 
bias of the gas is due to the fact that smaller, and therefore relatively 
more weakly clustered, halos have larger gas fractions then their larger 
and more strongly-clustered counterparts (see \S~\ref{sub:halo_properties}).  
At $k \gtrsim 10 \mpch$, the gas power spectrum rises dramatically, indicating that gas
has cooled and condensed into dense clumps of cold gas within dark
matter halo centers.  Both the dark matter and total matter power
spectra rise dramatically on scales $k \gtrsim 1 \mpch$.  These scales 
approximately correspond to the sizes of the largest halos in our
simulations ($M \sim$~a~few~$\times 10^{14} \hmsun$) and this dramatic
feature reflects the fact that gas has cooled and condensed in
halo centers causing a concomitant contraction of the dark matter
halos themselves 
\citep[{\it e.g.,}~][]{zeldovich_etal80,blumenthal_etal86,gnedin_etal04,sellwood_mcgaugh05}.
We elaborate on this point in \S~\ref{sub:halo_properties}.

Notice that the total matter power spectrum lies above both
the gas and dark matter power spectra.  The total matter power
spectrum includes a stellar component not depicted in Figure~\ref{fig:dpk}, 
which is significantly more 
clustered on these scales than either the gas or the dark matter.
Again, the effects of dissipation on the matter power spectra 
that we find agree qualitatively with the 
results of \citet{jing_etal06}. However, we find a larger relative
enhancement in both the dark matter and total matter power spectra in
the cooling case than is reported by \citet{jing_etal06}.  Also, we
find a sharp increase in the gas power spectrum at $k \gtrsim 10
\mpch$ in the $\csf$ simulation which is not seen in the corresponding
simulation of \citet{jing_etal06}.  The source of these differences is not entirely clear, but 
could be attributable to the specifics of the implementations of cooling, star 
formation, and feedback in the numerical codes used, to different gaseous equations 
of state or, at least in part, to sample variance.  
Finally, we note that the cooling
simulation exhibits a bias of $\sim 2-5\%$ in the 
total matter power spectrum 
at the largest scales ($k \lesssim 0.3 \mpch$).  
We are unable to explain this 
bias quantitatively and it is observed neither in 
our non-radiative simulation nor the simulation 
of \citet{jing_etal06}.  Finite volume along with the different 
nonlinear evolution of the simulation with baryonic condensation are 
likely culprits for the small differences in clustering on large scales.

\subsubsection{Convergence Power Spectra}
\label{subsub:powerconvergence}

While it is simplest to examine changes to the three-dimensional
power spectrum in the simulations, weak lensing surveys will 
measure the convergence power spectrum, which is a projection of 
$P(k)$ convolved with a lensing weight function.  Assuming a flat 
cosmological model and using Limber's approximation \citep[]{limber54,kaiser92}, 
the convergence power spectrum is given by 
\begin{equation}
\label{eq:Pkappa}
\Delta^2_{\kappa}(\ell) = 
\frac{\pi(\ell+1)}{\ell^2}\int_0^{z_s} dz\ 
\frac{W^2(\chi)\chi}{H(z)}\ \Delta^2(\ell/\chi,z)
\end{equation}
where $\Delta^2(k,z) \equiv k^3P(k,z)/2\pi^2$, 
the lensing weight is 
$W(z) = 3 \Omega_{\mathrm{M}} H_0^2 g(\chi)(1+z)/2$, 
and $\chi$ is the comoving distance to redshift $z$.  The function 
$g(\chi) = \chi \int_\chi^\infty d\chi' n(\chi) (\chi'-\chi)/\chi'$, 
where $n(\chi)$ describes the distribution of lensed 
source galaxies normalized so the integral over all redshifts 
is unity, $\int n(\chi) \ d\chi \ =\ 1$.  The distribution of
source galaxies is typically defined per redshift interval,
and we adopt the notation $n(z)$ as no confusion should arise.

We illustrate the 
importance of the effects of baryonic physics 
on the convergence power spectrum as follows.  
We assume for simplicity that the source distribution 
is given by a thin sheet of sources at $z_s=1$, 
so that $n(z) = \delta(z-z_s)$.  
Denoting $\chi_s(z_s)$ as the comoving distance to 
redshift $z=z_s=1$, this assumption gives 
$g(\chi) = \chi(\chi_s-\chi)/\chi_s$.  
Using more common source galaxy redshift distributions, 
such as $n(z)\propto z^2\exp(-[z/z_0]^2)$ as in 
\citet{ma_etal06} or $n(z) \propto z^2\exp(-[z/z_0])$ as 
in \citet[][$z_0 \sim 1$ in both cases]{huterer02}, 
affects the relative convergence power 
spectra that we quote at a level that is small compared to 
the baryonic processes that are the focus of this paper.  
We adopt a thin sheet of sources for simplicity as 
this choice makes our relative convergence spectra 
directly comparable to those presented by \citet{jing_etal06}, 
yet it remains a good representation of the relative spectra 
with more common source galaxy redshift distributions.
We compute the integral in Eq.~(\ref{eq:Pkappa}) 
by tabulating $\Delta^2(k,z)$ on a grid in the 
space of wavenumber and redshift using the outputs 
of the three simulations from $z=0$ to $z=1$.  
There are $41$ outputs available 
from the $N$-body simulation, 
$30$ outputs available from the 
non-radiative gas simulation, and 
$34$ outputs from the 
simulation with cooling and star 
formation.  We extrapolate outside of the 
range of wavenumbers probed by our simulations 
using the halo model, the details of which 
are described in \S~\ref{section:halomodel}.

We compare this result to the statistical 
uncertainty in the measurement of the convergence power 
spectrum for both contemporary and long-term, future
weak lensing surveys.  We include both sample variance and 
intrinsic shape noise 
under the assumption of a Gaussian 
density field \citep[{\it e.g.,}~][]{kaiser98}, 
\begin{equation}
\label{eq:wlerror}
\frac{\sigma_{\Delta^{2}_{\kappa}}}{\Delta^2_{\kappa}} = 
\sqrt{\frac{2}{f_{\mathrm{sky}}(2\ell + 1)}}
\Bigg[1 + \frac{\ell^2 \gamma^2}{2 \pi \bar{n} \Delta^2_{\kappa}}\Bigg], 
\end{equation}
where $f_{\mathrm{sky}}$ is the fraction of 
sky coverage of the survey, $\bar{n}$ is the effective number density 
of source galaxies on the sky, and $\gamma^2$ is the {\tt rms} 
intrinsic galaxy ellipticity.  Note that the error assumes 
Gaussian statistics whereas non-Gaussian effects are likely 
to be considerable over a large part of the range of scales where baryonic 
effects are also important \citep[][]{cooray_hu01,semboloni_etal06}.
As a near-term observational program, we consider a survey 
like the Canada-France-Hawaii-Telescope (CFHT) Legacy 
Wide Survey\footnote{{\tt http://www.cfht.hawaii.edu/Science/CFHLS/}} 
covering a fraction $f_{\mathrm{sky}}=4 \times 10^{-3}$
of the sky, with a number density of source galaxies of 
$\bar{n}=13 \ \mathrm{arcmin}^{-2}$.  We also consider the 
relative importance of baryonic physics compared to 
future weak lensing surveys such as LSST and SNAP.
For concreteness, we take $f_{\mathrm{sky}}=0.5$ and 
$\bar{n}=50 \ \mathrm{arcmin}^{-2}$ for our fiducial LSST-like 
survey and $f_{\mathrm{sky}}=0.025$ and $\bar{n}=100 \ \mathrm{arcmin}^{-2}$ 
for our fiducial SNAP-like survey.  
We adopt $\gamma^2=0.22$ for the variance in intrinsic source 
galaxy ellipticity and a bandwidth of $\Delta \ell/\ell = 1/10$.  
We emphasize that these are estimates of statistical 
uncertainties under the assumption of Gaussian statistics 
and include neither deviations from Gaussianity nor 
systematic uncertainties.

\begin{figure}
\plotone{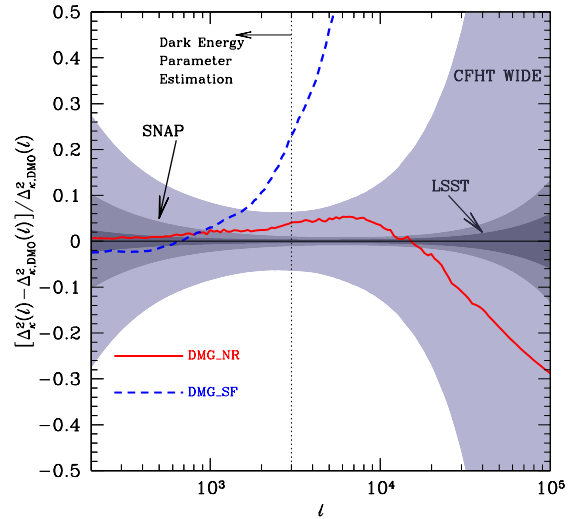}
\caption{ Fractional difference of the convergence power spectra
derived from the non-radiative $\ad$ ({\it solid line\/}) and cooling
$\csf$ ({\it dashed line\/}) simulations with gasdynamics relative to
the power spectrum of dissipationless $N$-body simulation $\nbody$.
The three shaded bands correspond to the statistical errors expected
from the CFHT Wide Survey, SNAP, and LSST (from outermost to innermost
band, respectively).  Note that these error bands are based on the
assumption of Gaussian statistics and do not include systematic
uncertainties.  Scales left of the dotted vertical line at $\ell=3000$ 
are those scales typically used to forecast constraints on dark energy parameters 
from future weak lensing surveys.  This demarcation is meant only as an 
approximate guideline.}
\label{fig:dcl}
\end{figure}

The resulting fractional difference of the convergence power spectra 
from that of the dissipationless $N$-body simulation are 
shown in Figure~\ref{fig:dcl}.  The relative difference in 
the convergence power spectrum in the cooling simulation is as dramatic
as was seen for $P(k)$.  For both the contemporary CFHT Wide 
Survey and the future LSST- and SNAP-like surveys, the 
systematic difference is far greater than statistical 
uncertainties for $\ell \gtrsim 10^{3}$.  In practice, 
dark energy constraints are derived using multipoles 
less than some $\ell_{\mathrm{max}}$ where deviations 
from Gaussianity are assumed to be unimportant 
\citep[{\it e.g.,}~][]{white_hu00,vale_white03}.  The 
vertical line at $\ell = 3000$ in Fig.~\ref{fig:dcl} 
represents a typical value of $\ell_{\mathrm{max}}$ and 
is meant to serve as a rough guideline for scales relevant 
to forecasts that appear in the literature; however, a wide 
range of choices for $\ell_{\mathrm{max}}$ have been made 
\citep[{\it e.g.,}~][]{huterer02,refregier03,refregier_etal04,ma_etal06,zhan06}.

One might argue that current hydrodynamical simulations are 
not yet up to the task of predicting accurate 
power spectra due to notable shortcomings in modeling 
galaxy formation, such as 
the well-documented ``overcooling'' problem 
\citep[{\it e.g.,}~][]{katz_white93,suginohara_ostriker98,lewis_etal00,
pearce_etal00,dave_etal01,balogh_etal01,borgani_etal02}.
However, notice that even in the case of the non-radiative 
simulation the power spectrum deviates from the $N$-body 
case at a level that will be important to future 
efforts such as SNAP and LSST.  This result suggests that our 
inability to make robust predictions for the evolution of 
the baryonic component of the Universe introduces a large, 
systematic uncertainty into the theoretical predictions 
for the growth of perturbations, and ultimately for 
the analysis and interpretation of such data.

\subsection{Simulated Halo Properties}
\label{sub:halo_properties}

Although density perturbations are small and grow according to linear theory on
large scales, on the small scales probed by weak lensing surveys 
typical density fluctuations approach or exceed unity and fluctuations 
on these scales evolve nonlinearly.  
In fact, the power spectrum at $k \gtrsim 0.5 \mpch$ is dominated by the 
structures of virialized dark matter halos
\citep[{\it e.g.,}~][]{scherrer_bertschinger91,ma_fry00,zhan_knox04}.  
Given that the differences in the power spectra discussed in \S~\ref{sub:spectra}
are primarily confined to these scales, we find it instructive to
explore changes within these highly-nonlinear, bound objects.  In this
section we present the systematic differences in halo properties
between the $\nbody$, $\ad$, and $\csf$ simulations.

\subsubsection{Halo Mass Functions and Gas Fractions}

\begin{figure}
\plotone{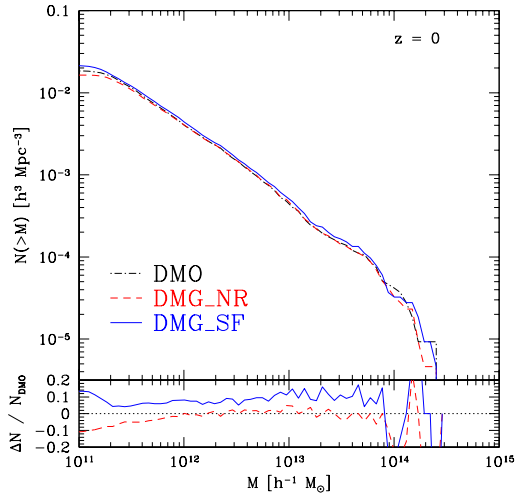}
\caption{
{\it Upper Panel}: Cumulative mass 
function of dark matter halos.   We plot the 
number of halos with virial mass greater than 
$M$ as a function of $M$. 
{\it Lower Panel}: 
The fractional deviation from the dissipationless
mass function.  We plot the difference between 
the number of halos in the hydrodynamical simulations 
and the $N$-body simulation normalized by the 
number of halos in the $N$-body simulation, 
$[N(>M)-N_{\nbody}(>M)]/N_{\nbody}(>M)$.
\label{fig:mf}}
\end{figure}

We begin with the abundance of halos as a function of halo mass.  
Figure~\ref{fig:mf} shows the cumulative mass functions 
for the three simulations.  The mass functions in 
the $N$-body simulation and the non-radiative simulation 
are very similar for masses above $\sim 10^{12} \hmsun$.  
The two mass functions differ at lower masses
due to differences in numerical resolution of the 
$\nbody$ and $\ad$ runs. Consequently, 
we only consider halos with virial masses greater than
$10^{12} \hmsun$ ($\approx 1000$ dark matter particles) when
comparing halo properties.

The cumulative mass function in the $\csf$ simulation is 
consistently $\sim 10\%$ higher than in the $\nbody$ 
simulation over a wide range in mass.  The strong redistribution
of mass in the $\csf$ halos (see \S~\ref{sub:profiles}) causes the 
halo mass (defined at a fixed overdensity) to be larger for the 
same halo between the $\nbody$ and $\csf$ simulations, effectively 
shifting the cumulative mass function to larger masses.  The 
clustering of these halos remains unchanged, however, which leads
to a significant change in the halo bias-mass relation, a rapidly
varying function of mass on cluster scales.  

\begin{figure}
\plotone{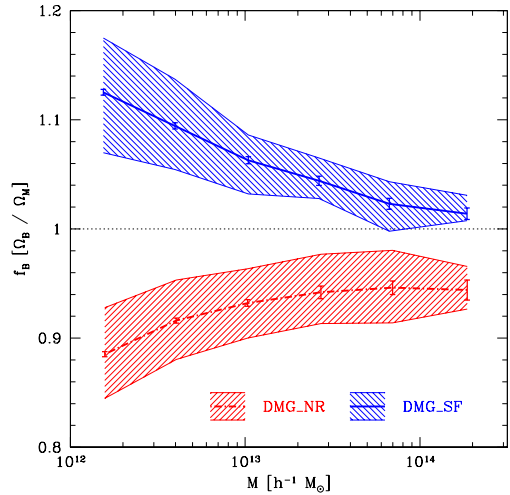}
\caption{ The mean fraction of mass in baryons within the virial
radius as a function of halo mass for halos in both the non-radiative
simulation ({\it dash-dotted curve}) and the cooling+star formation
simulation ({\it solid curve}).  The baryon fraction is expressed in
units of the universal value assumed in our simulations
$\Omega_{\mathrm{B}}/\Omega_{\mathrm{M}}$.  Note that the halo baryon
fractions in the non-radiative simulation are below the universal
value (exhibiting the so-called baryon bias), while the baryon
fractions in the simulation with cooling and star formation are above
the universal value.  Hashed regions correspond to the 68\% intrinsic
scatter amongst halos, while error bars correspond to the estimated
error of the mean.\label{fig:fbaryon} }
\end{figure}

In the two hydrodynamic simulations ($\ad$ and $\csf$), we can also
track the manner in which baryons are apportioned to dark matter
halos.  Figure~\ref{fig:fbaryon} shows the baryon fraction as a
function of halo mass normalized to the universal baryon fraction
($\Omega_{\mathrm{B}}/\Omega_{\mathrm{M}} = 0.143$ in our cosmology).
The halos in the non-radiative simulation exhibit baryon fractions
that are $\approx 5-10\%$ below the universal value, consistent with
previous simulations without cooling
\citep[]{eke_etal98,frenk_etal99,ettori_etal06,gottloeber_etal06,crain_etal06}.
Infalling baryons are heated by shocks at distances of order the
virial radius or larger, exchange energy with the dark matter
component during relaxation, and thus no longer follow the same
distribution as the dark matter.  Instead, the gas follows a more
diffuse, extended profile and never achieves extremely high densities
in halo centers.  Rather, as we discuss below, the gas profiles in the
$\ad$ simulation exhibit cores of nearly constant density that extend
to large fractions of halo virial radii.  Note that the slight decline
in the halo mass-baryon fraction relation toward $M_{\mathrm{vir}} \sim 10^{12}\ h^{-1}M_{\odot}$
may be due to limited numerical resolution.  We leave to future studies the question
of whether the mass-baryon fraction relation truly varies with mass
in the non-radiative regime or whether there exists a fixed baryon-bias
which is constant with mass.

Halos in the $\csf$ simulation, on the other hand, have a $\sim 5-10\%$ excess of 
baryons relative to the universal value.  This excess is caused by the ability of 
gas to dissipate kinetic energy and condense at the center.  
These baryons are then unlikely to make large excursions 
away from the halo center and they make the potential well seen by newly 
accreted material considerably deeper than in the $N$-body or 
non-radiative cases.

\subsubsection{Halo Density Profiles}
\label{sub:profiles}

Halos in dissipationless $N$-body simulations have 
spherically-averaged density profiles that are
well described the profile of \citet[][NFW]{navarro_etal97}, 
\begin{equation}
\label{eq:nfw}
\rho_{\mathrm{NFW}}(r) \propto \frac{1}{(c r/\Rvir)(1+c r/\Rvir)^2}.
\end{equation}
The relative concentration of mass toward the halo center is 
described by the concentration parameter $c$.  The concentration 
parameter has been studied by numerous authors and 
the distribution of concentration parameters of halos 
at fixed mass is known to be a function of halo 
mass \citep[{\it e.g.,}~][]{navarro_etal97,bullock_etal01}.  

The baryonic component in our $\ad$ simulation follows
the distribution of dark matter at large scales ($\approx r^{-3}$);
however, at small scales the gas density profile 
tends toward a constant density core that
is well described by the profile of \citet[]{burkert95},
\begin{equation}
\label{eq:burkert}
\rho_{\mathrm{B}}(r) \propto
\frac{1}{(1 + c_{\mathrm{B}}r/\Rvir)(1 + [c_{\mathrm{B}}r/\Rvir]^2)}.
\end{equation}

We compute the spherically-averaged density profile 
for each of the halos in our three simulations by 
binning the cumulative mass profile in 100 spherical 
bins logarithmically spaced in radius from $5-2000 \hkpc$ 
centered on the dark matter particle with the highest 
local density. Dark matter and stellar particles are 
treated as point masses, while gas is assigned to bins 
by numerically integrating the gas density within radial 
annuli using $10^5$ Monte Carlo samples per bin.

We obtain fitted concentrations for each halo by minimizing the $\chi^2$
statistic with bins weighted by the square-root of the bin density.  In the case 
of particles (dark matter and stellar mass) this procedure is equivalent to assuming the 
bin error follows a Poisson distribution.  Profiles are fit from an innermost
radius $R_{\mathrm{min}} = 0.05 \Rvir$ to $\Rvir$.  In practice the radii used
are approximate due to the fixed widths of the bins in physical radius.  Fits to 
the halos in the $\nbody$ and $\ad$ simulations are insensitive to the choice of innermost
radius, but the density profiles in halos in the $\csf$ simulation deviate
strongly from the NFW profile within this radius, leading to concentration
fits that are biased to larger values as $R_{\mathrm{min}}$ is decreased.  
The deviation is caused by the formation of a large, central galaxy in the 
$\csf$ case.  In addition to fitting the total matter distributions 
and the dark matter distributions for all three simulations, we fit the gas distributions 
inside the halos of the non-radiative simulation 
with the Burkert profile [Eq.~(\ref{eq:burkert})] to obtain a Burkert concentration 
parameter $c_{\mathrm{B}}$.

\begin{figure*}[t]
\centerline
{
	\vspace*{2pt}\epsfysize=2.25truein \epsffile{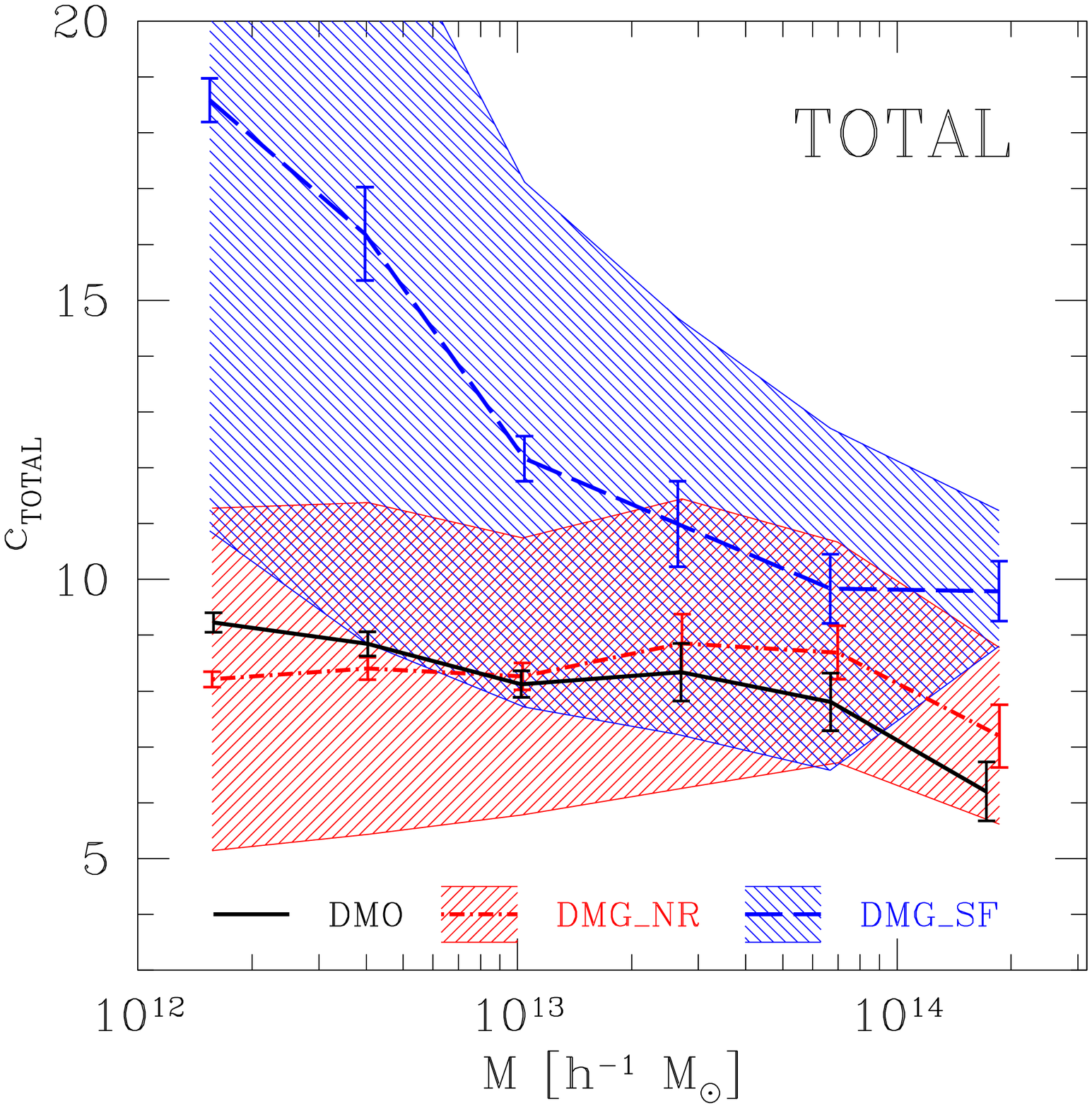}
	\hspace*{2pt}\epsfysize=2.25truein \epsffile{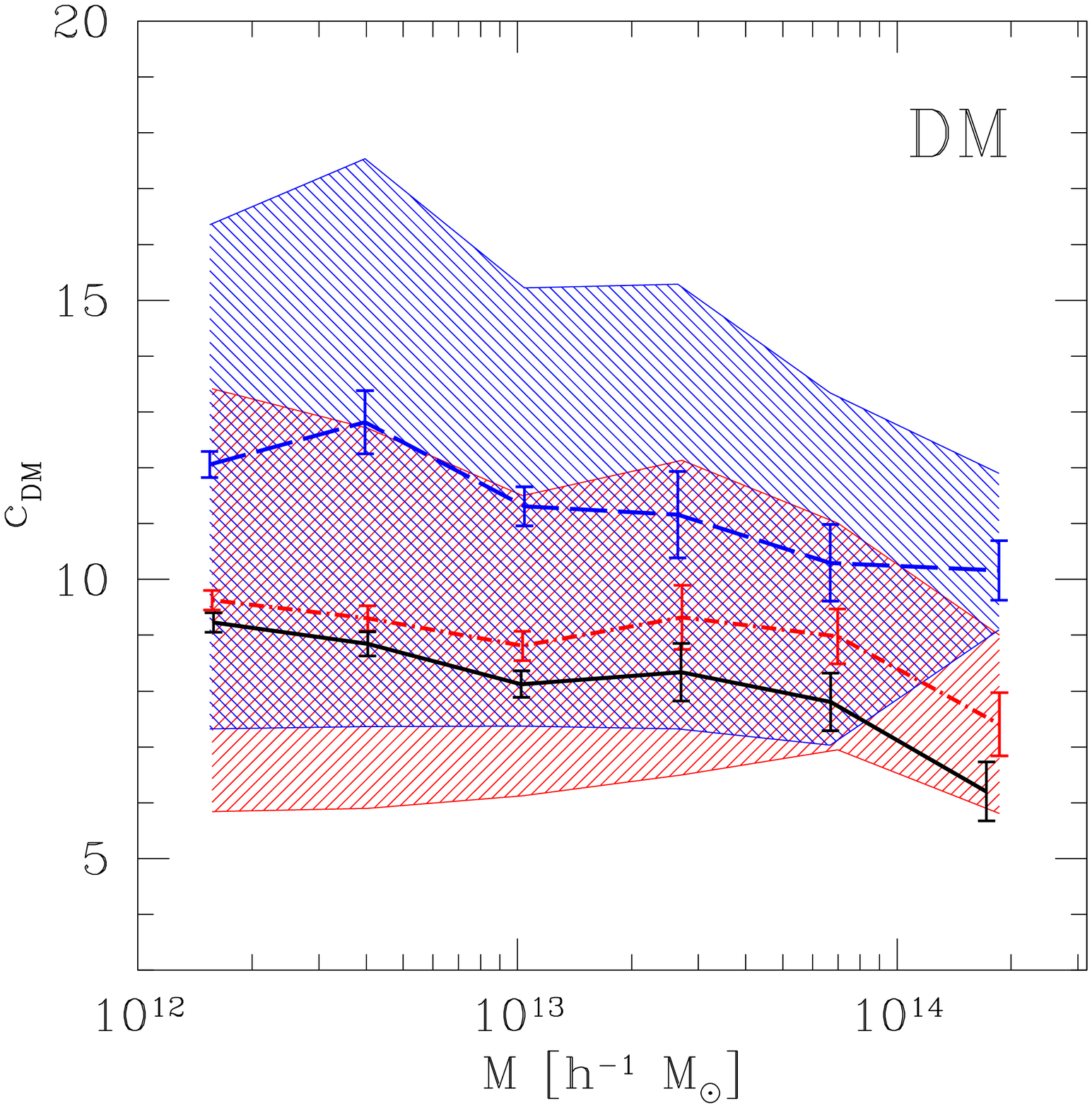}
	\hspace*{2pt}\epsfysize=2.25truein \epsffile{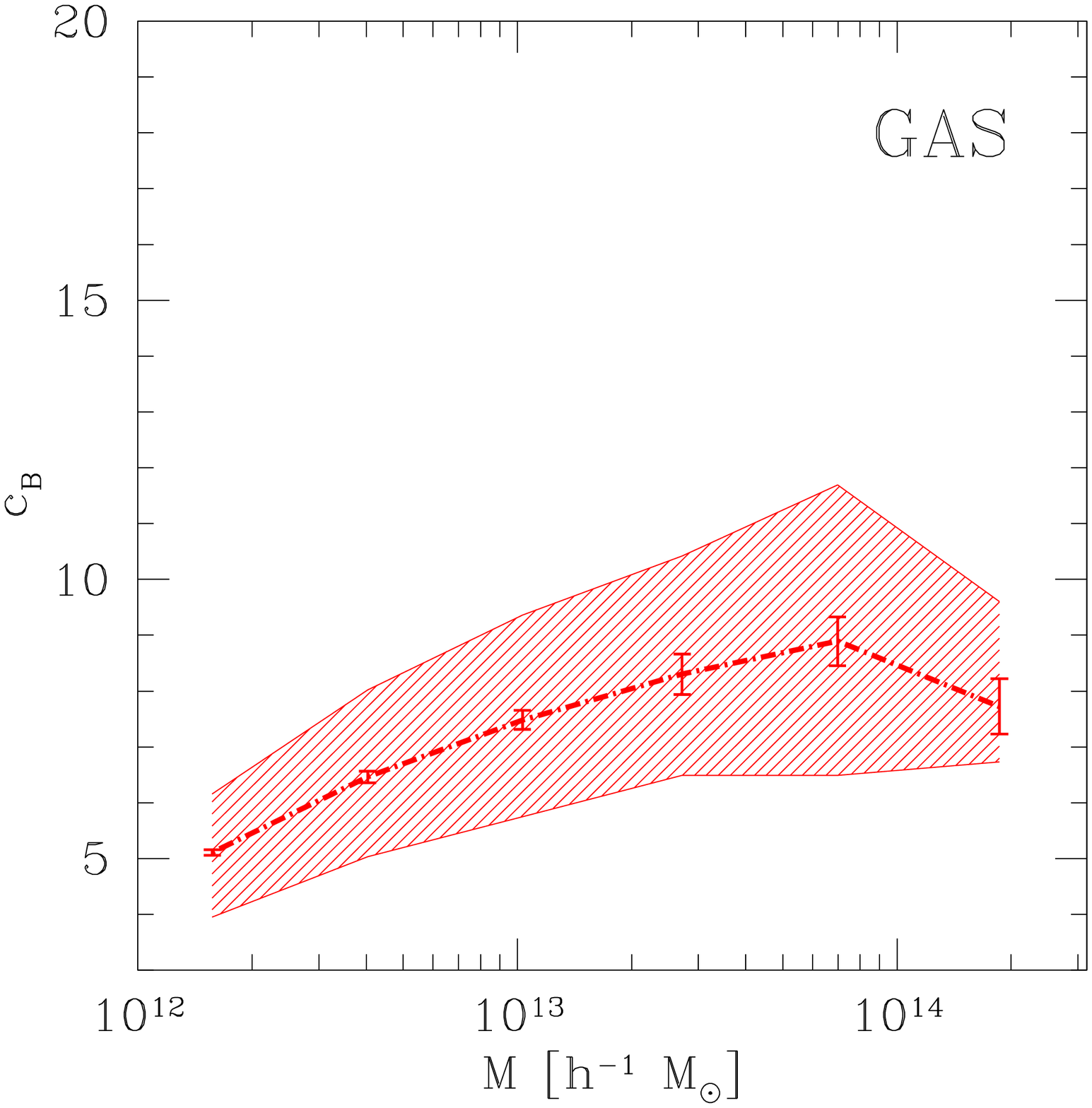}
}
\caption{Mean concentration of radial distributions of main mass
components as a function of halo mass. Shaded regions contain 68\% of
halos in a given mass bin.  Scatter in the concentration relation for
the $\nbody$ simulation is comparable to that in the $\ad$ simulation
and is omitted for clarity.  Error bars correspond to the estimated
error on the mean concentration in each bin.  {\it Left panel}: the
mean NFW concentration fit to the {\it total} (i.e., DM$+$gas$+$stars)
mass distribution for all three of our simulations as indicated at the
bottom of the panel. {\it Center panel}: the mean NFW concentration
fit to the {\it dark matter\/} distribution in all three
simulations. {\it Right panel}: the mean Burkert concentration fit to
the gas density profiles in the $\ad$ simulation.  }
\label{fig:cm}
\end{figure*}

Figure ~\ref{fig:cm} shows the resulting concentration parameters as a function 
of halo mass derived from these density profile fits.
Several important features are apparent.  First, consider the dark matter 
concentrations in the center panel.  The halos in the $\ad$ simulation 
are slightly more concentrated ($\sim 10\%$) than the halos in the $\nbody$ 
simulation, which is consistent with the effects found by previous simulation
studies \citep{rasia_etal04,lin_etal06}.  As discussed earlier, during halo 
formation and relaxation the gas exchanges energy with the dark matter.  The 
net result is that the gas gains energy while the dark matter relinquishes energy 
to the gas, resulting in slightly more compact dark matter halos and extended 
gas distributions.  

Furthermore, Figure~\ref{fig:cm} shows that the halos of the $\csf$
simulation are considerably more concentrated than the halos of both
the $\nbody$ and $\ad$ runs.  This difference is due to the
cooling-induced condensation of gas and formation of galaxies at the
centers of these halos.  The left panel of Figure~\ref{fig:cm} shows the
NFW concentrations that result from fitting the total matter
distributions of the halos in each simulation.  The results are
qualitatively similar to those of the dark matter distributions alone,
but exhibit two additional features for halos with masses $\lesssim
10^{13} \hmsun$.  The first feature is a relative decline in total
matter concentration in the $\ad$ simulation relative to the $\nbody$
simulation.  This decline is due to the extended, less-concentrated
gas distribution in these lower-mass halos, and is again consistent
with \citet{lin_etal06} who find a smaller ($\sim 3\%$) increase in
the concentration of the total matter distribution than for the dark
matter itself. The second feature for low-mass halos is a dramatic
rise in total matter concentration in the $\csf$ simulation.  This
increase in concentration is because halos with masses $\lesssim
10^{13} \hmsun$ cool gas and form galaxies more efficiently than their cluster-sized
counterparts.  The larger fraction of cold gas and stars at the
centers of these halos gives rise to a large relative increase in
density toward the halo centers, driving the concentration values higher.
Finally, the right panel of Figure~\ref{fig:cm} gives the Burkert
concentrations of the gas profiles in the $\ad$ simulation.  Note the
gradual increase in $c_{\mathrm{B}}$ with halo mass.  This result will
be discussed further in \S~\ref{sub:halomodelmethods}.

\section{Interpreting Power Spectra with the Halo Model}
\label{section:halomodel}

\subsection{Halo Model Background and Methods}
\label{sub:halomodelmethods}

To interpret the relative differences seen in the simulated power spectra
(Figure~\ref{fig:dpk}), we turn to the halo model --- an analytic, phenomenological framework 
for describing the clustering of dark matter, galaxies, or any 
other population associated with dark matter halos.
Various aspects of the halo model are developed and 
discussed in a variety of studies over several decades 
\citep{neyman_scott52,peebles74,mcclelland_silk77,scherrer_bertschinger91,ma_fry00,seljak00,scoccimarro_etal01,sheth_etal01b,sheth_etal01a,berlind_weinberg02,cooray_sheth02}.

The halo model is predicated on the assumption that all of the objects 
of interest lie within dark matter halos.  
The two-point clustering statistics of matter are then given by the the sum of 
two terms.  The ``one-halo'' term is from matter elements residing within a common halo 
and dominates clustering statistics on scales smaller than typical halo sizes. 
The second ``two-halo'' term is from matter elements residing within 
two distinct halos and dominates on large scales.  
This decomposition is convenient because we do not expect baryonic physics  
to alter significantly the clustering of dark
matter halos on large scales.  Baryons do have significant
effects on the structures of individual dark matter halos, as shown
in \S~\ref{sub:halo_properties}, so we expect differences to be 
largely confined to the one-halo term.

To model the total matter clustering with 
multiple components (e.g., dark matter, gas, etc.), it is 
convenient to treat each component separately, as was done by
\citet{zhan_knox04}.  In what follows, we quote the halo model
relations for the power spectrum.  
The total power is the sum of the power spectra
of the individual components and cross terms,
%
%
$P(k) = \sum_{i,j} f_i f_j P_{ij}(k)$, 
%
%
where $f_i = \Omega_i/\Omegam$ refers to the universal 
mass fraction in the $i$th matter component.
The one-halo contribution is given by
\begin{equation}
\label{eq:P1h}
P^{1\mathrm{H}}_{ij}(k) = \frac{1}{\rho_i \rho_j} 
\int \dd m\,m^2 f_i(m) f_j(m) \frac{\dd n}{\dd m} \lambda_i(k;m) \lambda_j(k;m),
\end{equation}
where $\rho_i$ is the mean density in the $i$th matter component, 
$f_i(m)$ is the average fraction of mass in halos of total mass 
$m$ residing in the $i$th component, 
$\dd n/\dd m$ is the mass function of halos, 
and $\lambda_i(k;m)$ is the Fourier transform of 
the mean density profile of the $i$th component in 
halos of total mass $m$.  For example, the profiles of 
dark matter halos are often modeled by NFW profiles, in 
which case, $\lambda_i(k;m)$ is the Fourier transform 
of the NFW density profile \citep[e.g., given by][]{scoccimarro_etal01} with 
a concentration parameter set by some relation \citep[{\it e.g.,}~][]{bullock_etal01}.  
The two-halo contribution to $P(k)$ is
\begin{equation}
\label{eq:P2h}
P^{2\mathrm{H}}_{ij}(k)  =  \frac{1}{\rho_i \rho_j} P^{\mathrm{lin}}(k) B_i(k) B_j(k),
\end{equation}
where
\begin{equation}
\label{eq:bdef}
B_i(k) \equiv \int \dd m\,m f_i(m) \frac{\dd n}{\dd m} \lambda_i(k;m) b_\mathrm{h}(m),
\end{equation}
$P^{\mathrm{lin}}(k)$ is the linear matter power spectrum, 
and $b_\mathrm{h}(m)$ is the mass-dependent halo bias.

Our primary aim in applying the halo model is to study 
the qualitative features of the spectra from our simulations 
rather than to provide a precise, quantitative description.  
Therefore, we adopt the fitting forms for the mass function 
and linear bias of dark matter halos 
provided by \citet{sheth_tormen99}, rather than any of several updated bias 
prescriptions 
\citep[{\it e.g.,}~][see \citealt{cooray_sheth02} and \citealt{zentner06} recent reviews]{jenkins_etal01,seljak_warren04}.  
This choice guarantees that the two normalization relations 
\begin{equation}
\frac{1}{\rho} \int\ \dd m \ \frac{\dd n}{\dd m} = 1
\end{equation}
and
\begin{equation}
\int\ \dd m\ \frac{\dd n}{\dd m} \Bigg(\frac{m}{\rho}\Bigg) b_{\mathrm{h}}(m) = 1
\end{equation}
are satisfied identically without making any further, and often arbitrary, 
choices about how these relations should be enforced.  
As halos have a finite extent set by their virial 
radii, the integrals in Eq.~(\ref{eq:bdef}) should not extend over all 
mass but should be limited to halos with virial radii smaller than 
$r \sim k^{-1}$.  This effect is known as halo exclusion.  Though more 
complex and accurate implementations of halo exclusion exist 
\citep[{\it e.g.,}~][]{tinker_etal06}, we use the model for halo exclusion introduced by \citet{zheng03}. 
Briefly, we set the upper bounds on the integrals in Eq.~(\ref{eq:P2h})
to the halo mass that corresponds to a virial
radius of $r_{\mathrm{max}} = 2\pi k^{-1}$.
Previous studies have found this prescription to be 
useful for practical applications \citep[{\it e.g.,}~][]{zheng03,zehavi_etal04}.

The last ingredients necessary to build a halo model of the matter power
spectrum are specifications of the density profiles that characterize the distribution 
of each matter component within halos.  We treat each of the cases of pure dark matter, 
dark matter with non-radiative gas, and dark matter with gas cooling 
and star formation slightly differently, with 
prescriptions motivated by our set of simulations.

We model the dark matter halos in both the $N$-body and non-radiative
cases with the NFW density profile [Eq.~(\ref{eq:nfw})].  As in 
\S~\ref{sub:profiles}, the concentrations of halos are 
different in each case, and we include this effect in our
implementation of the halo model.  In our modeling, it is necessary to
extrapolate beyond the range of concentrations probed directly by our
simulations.  Partly motivated by the fact that we aim to represent
the features of our simulated spectra qualitatively, we adopt a
particular form of the analytic model for halo concentrations
introduced in \citet{bullock_etal01}.  Similar to other authors
\citep[{\it e.g.,}~][]{dolag_etal04,kuhlen_etal05,wechsler_etal06,maccio_etal06}, 
we find that the relationship between concentration and mass in our
simulations has a smaller normalization and a slightly shallower slope
than that of the \citet{bullock_etal01} model in its original form.
We find that the {\em mean} concentration as a function of mass in the $\nbody$ simulation is
well described by the \citet{bullock_etal01} model with parameters
$F=10^{-5}$ and $K=1.7$.  We stress that these parameters are {\em not} the
result of a formal fitting procedure and defer further exploration of
the concentration-mass relation to future work.

The halos in our $\ad$ simulation exhibit somewhat higher
concentrations than those in the $\nbody$ simulation.  
Over the mass range measured in the
simulation, we use the measured concentration-mass relation (e.g.,
Fig.~\ref{fig:cm}) from the simulations 
and use the \citet{bullock_etal01}-like model in order to
extrapolate outside of the range of measured halo concentrations.
Specifically, we fix the \citet{bullock_etal01} parameter $F=10^{-5}$
as for the dissipationless simulation and allow $K$ to float in order
to best match the normalization of the concentration-redshift relation
at the redshift of each simulation output.  Typical values of $K$ 
are $\sim 10\%$ larger than in the dissipationless case.

Including gas in the halo model for the non-radiative simulation introduces
additional complexity into the standard halo model.  It is now
necessary to specify how much gas there is inside each halo.  
Figure~\ref{fig:fbaryon} gives the mean fraction of halo mass in baryons 
as a function of halo mass for both simulations with baryons.  
In the non-radiative case we adopt a constant value of 
$f_{\mathrm{B}} = 0.94 \Omega_{\mathrm{B}}/\Omega_{\mathrm{M}}$ 
for all halos.  This value is consistent with the baryon fractions in the largest halos of 
$\ad$ to redshifts greater than $z=3$.  Treating the baryon fraction in this way 
is sensible because the one-halo contribution to the 3D and 
convergence power spectra should be dominated by the largest halos 
on scales of interest (see Fig.~4 of \citet{zhan_knox04}) 
and insures that we do not model a resolution effect.  For simplicity, 
we treat the remaining baryons as an unclustered component.

Our non-radiative simulation revealed that the NFW profile is not a good 
description of the distribution of gas within halos (\S~\ref{sub:profiles}). In this
case we model the gas separately with a Burkert profile.  The constant-density core 
within $r \ll \Rvir/\cburk$ causes a suppression in the small-scale power at high wavenumbers 
relative to an NFW profile.  The trend of $\cburk$ as a 
function of halo mass defined in this way 
is the opposite of the NFW concentrations of halos.  
An exploration of the possible source of this
trend is beyond the scope of this work.  
We interpolate the mean concentration relation in Figure \ref{fig:cm} linearly in 
$\log(c_B)$ as a function of $\log(m)$.  For
masses below our resolution limit we assume a constant $c_B$ equal to that of the lowest mass
bin, although we have explicitly checked that our results are insensitive to this choice.
That our choice of extrapolation does not affect our results significantly is
unsurprising, as halos below $10^{12} \hmsun$ contribute only a small 
fraction of the power on scales of interest.  Likewise, we set $c_B$ equal to the value 
at the highest mass bin for halos larger than the halos in 
our simulation volume.

In principle, modeling power spectra in the case where radiative
cooling and star formation are included is more complex still.  The
halos in our simulations with radiative cooling and star formation
exhibit baryon density profiles that are significantly more complex
than those in the non-radiative simulations.  Instead of a single, hot
baryonic component, cooling and star formation give rise to a
condensed component of cold gas at the center of each halo, a fraction
of which turns into stars.  Moreover, the baryons in the gas phase
separate into hot and cold components that do not follow a common
density profile.

Describing all three components of the baryons separately 
would be difficult and, while interesting, we relegate this to 
future work focusing on galaxy formation physics.  
Fortunately, such a 
detailed exploration of the distribution of baryons within 
halos is not necessary in order to understand the 
qualitative features in the relative power spectra.
The process that causes the 
strong condensation of baryons at halo centers
also affects the dark matter, and the total matter profile 
is well described by an NFW profile outside $\approx 0.05-0.1 \Rvir$.
As such, we model the total mass profiles of halos in the case of 
radiative cooling and star formation as a single component described 
by an NFW profile with a concentration set according to the tabulated mean 
concentration-mass relation extracted from the cooling simulations (left panel
of Fig.~\ref{fig:cm}).  
We extrapolate outside of the mass range probed by our simulations 
in the same manner as for dark matter halos in the non-radiative case.

\subsection{Halo Model Power Spectra}

\begin{figure}
\plotone{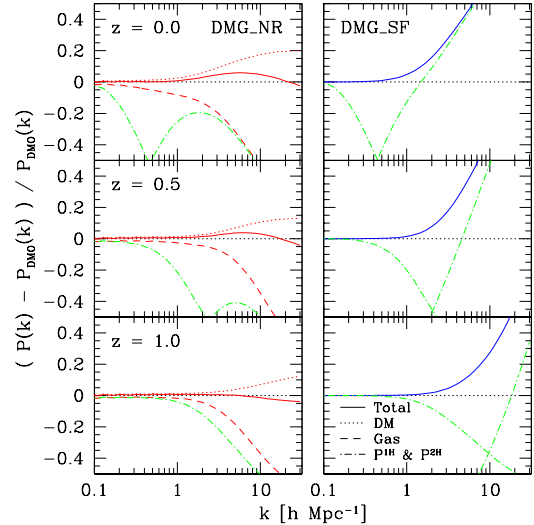}
\caption{The halo model power spectra for the non-radiative ($\ad$) and cooling+star formation
($\csf$) cases are plotted in left and right columns respectively.  Line types and panel layout
follow the convention of Figure~\ref{fig:dpk}.  The {\it dot-dashed} lines represent the one-
and two-halo contributions to the baryon power spectra (for the $\csf$ case baryons are assumed
to trace the total matter distribution).  The two-halo term dominates at low $k$, while the 
one-halo term is considerable at intermediate wavenumbers.  Note that the simple modifications to the 
standard halo model formalism described in the text reproduce the qualitative features of the 
simulated spectra well.
}
\label{fig:dpk_hm}
\end{figure}

The left-column of Figure~\ref{fig:dpk_hm} shows the power 
spectra from a halo model including the 
effects of non-radiative baryonic physics as described above.  At $z=0$ there is good
qualitative agreement between the halo model power spectra and the power 
spectra from the simulations.  
In particular, the qualitative features of an enhancement in the dark matter spectrum 
at high wavenumber ($k \gtrsim 1 \mpch$), a comparably suppressed gas power spectrum, 
and a total matter power spectrum  with an enhancement 
at intermediate wavenumber ($k \sim 1-10 \mpch$) and 
subsequent decrease at high wavenumber ($k \gtrsim 20 \mpch$) 
are all reproduced by this simple, augmented halo model.  

{\em All of these features are set by modifications to the
one-halo term in the halo model.}  The increase in the dark matter
power is caused by the increased concentrations of halos in the
non-radiative case.  The relative decline of the gas power spectrum is
due to the more extended profile of the gas relative to an NFW profile
[see Eq.~(\ref{eq:burkert})].  The two features in the total mass
power spectrum are a reflection of both of these effects.  

At higher redshifts these features move to larger wavenumbers, a shift
that is also present in the simulations but to a lesser degree.  In
the context of the halo model, this shift in scale is caused by the
change of the peak location of the one-halo term as the sizes of the
largest halos decrease with increasing redshift.  We attribute differences
between the redshift evolution of the simulations and of the halo model
to differences in the evolution of the halo mass function and bias relations 
in the simulation, which are not exactly given by the analytic \citet{sheth_tormen99}
relations.  In particular, the mass function exhibits large variations 
at the high-mass end due to the limited simulation volume.
There also exists an apparent
large-scale bias of the gas component because some fraction
of the gas in our simulations is not contained within halos and we have
treated this fraction as an unclustered component, whereas in the 
simulations these baryons are clustered, albeit somewhat less
strongly than the baryons inside halos.

The right-column of Figure~\ref{fig:dpk_hm} shows the halo power
spectra resulting from our model of the radiative case.  Again, the
qualitative features of the simulated spectra are reproduced by the
halo model.  The power increases steeply at $k \gtrsim 1\mpch$ at all
redshifts, although the transition is quicker at lower redshift,
reaching a $\sim 40\%$ increase of total power by $k \sim 4\mpch$ at
$z = 0$.  At higher redshifts this transition shifts to smaller
scales. Again, the relative importance of the one-halo term is responsible for
setting the scale that this occurs.  

The increase of power in the one-halo term is caused primarily by the
 larger halo concentrations in the $\csf$ run.  Increasing the 
 normalization of the $c(m)$ relation to the level observed in this
 simulation leads to a considerable enhancement of the matter power
 spectrum at small scales.  Finally, note that the slopes of the
 simulated power spectra continue to increase as a function of
 wavenumber at small scales, a feature not present in our halo model
 spectra.  This trend is likely due to the fact that our simple halo
 model is not designed to reflect the contributions of the condensed
 gas and stellar components within $r \lesssim 0.05-0.1 \Rvir$ of each
 halo, deviations from the NFW form, or halo substructure, 
and so it cannot represent the matter
 power spectra at very high wavenumbers well.

\begin{figure*}[t]
\centerline
{
        \epsfysize=2.80truein \epsffile{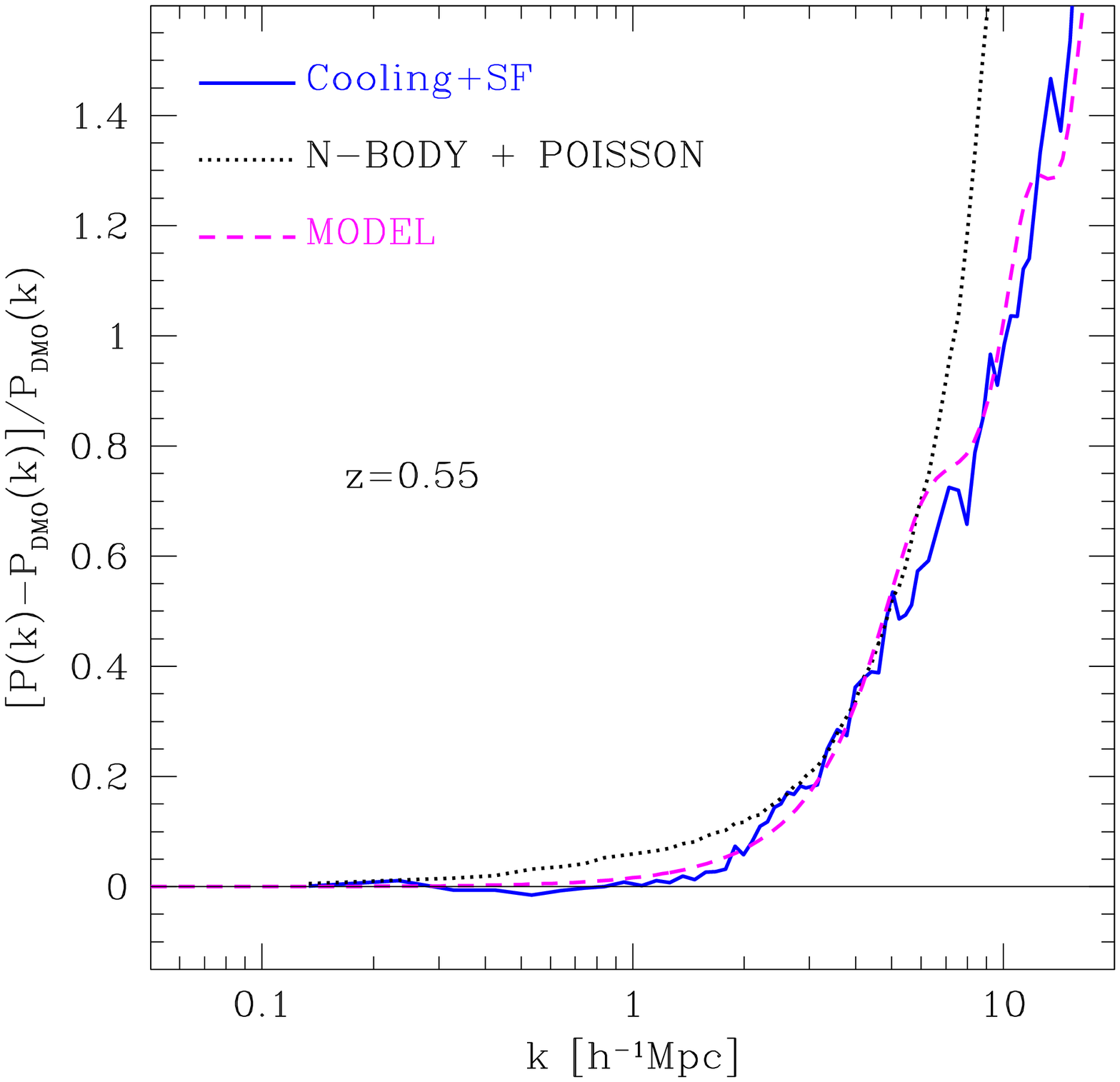}
        \hspace*{12pt}\epsfysize=2.80truein \epsffile{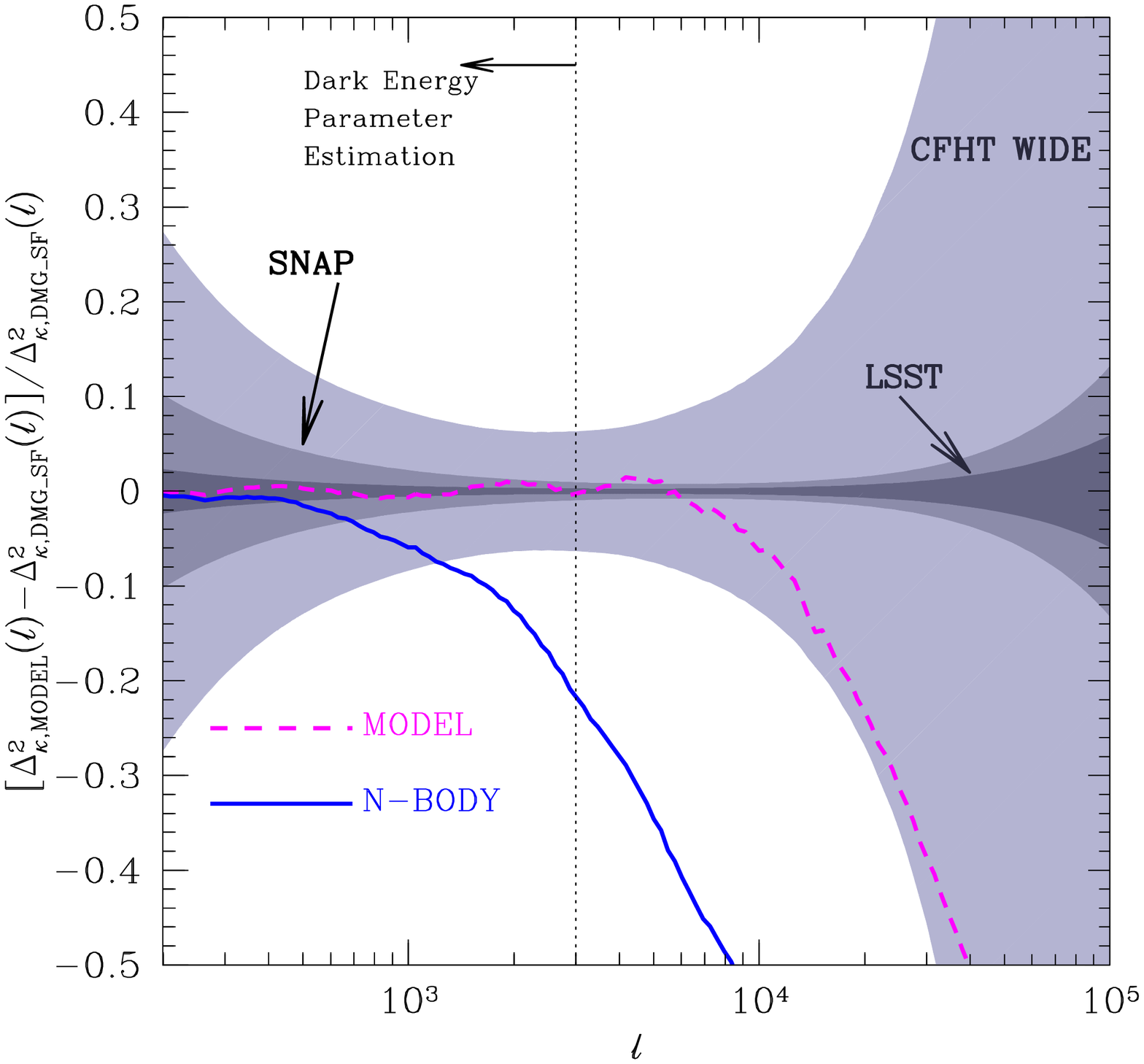}
}
\caption{
A model for the gross effect of baryon cooling on 
the matter power spectrum.  The {\em left} panel shows matter 
power spectra at $z=0.55$ (where the lensing weight function peaks 
given a thin plane of sources at redshift $z_s=1$), 
relative to the matter power spectrum 
in the $N$-body simulation.  In this case, we have scaled out 
the small, large-scale bias of the power in the cooling simulation relative 
to the $N$-body simulation at small wavenumber.  The {\em solid} 
line shows the result of the cooling simulation.  The {\em dotted} 
line shows an example of adding a Poisson term to the power spectrum 
of the $N$-body simulation.  The {\em dashed} line shows the result of 
taking our simple model for the net effect of baryonic dissipation on 
the matter power spectrum.  In 
this particular example, we take $\Rvir = 1.1 \hmpc$, $c_1=5$, 
and $c_2=1.7c_1=8.5$.  The {\em right} panel shows convergence power 
spectra relative to the power spectrum in the $\csf$ simulation.  
Note that this choice of normalization is different from Figure~\ref{fig:dcl},
which is why why the $\nbody$ case here appears as a deficit.
The line types are the same as in the left panel and the error bands 
are the same as those in Figure~\ref{fig:dcl}.  
}
\label{fig:model}
\end{figure*}

\subsection{An Heuristic Model For Power Spectra With Cooling and Star Formation}
\label{sub:hmodel}

The trends discussed in the previous subsections suggest that the primary effect of 
baryons is to redistribute matter toward the centers of halos.  In the case
where the baryon component is allowed to cool, this effect is particularly dramatic,
and can be adequately described over a wide range of scales by an increase in 
the effective NFW concentration (see \S~\ref{sub:profiles}).  Our
results indicate that the differences are largely relegated to the one-halo term
in the halo model [Eq.~(\ref{eq:P1h})].

This physical intuition suggests a simple method to model the effect of baryonic
cooling on the power spectrum over the limited range of wavenumbers
important for weak lensing studies.  The
Fourier transform of the NFW profile 
is analytic and given by \citep[{\it e.g.,}~][]{scoccimarro_etal01}
\begin{eqnarray}
\label{eq:lambdaknfw}
\lambda(\eta;c) & = & \frac{1}{f_{\mathrm{NFW}}(c)} \Bigg\{ \sin(\eta) [ \Si([1+c]\eta) - \Si(\eta)] \nonumber \\
 & & + \cos(\eta) [ \Ci([1+c]\eta) - \Ci(\eta)] \nonumber \\
 & & - \frac{\sin(\eta)}{[1+c]\eta} \Bigg\},
\end{eqnarray} 
where $c$ is the halo concentration, 
$\eta=kR_{\mathrm{vir}}/c$, $R_{\mathrm{vir}}$ is the
halo virial radius, $f_{\mathrm{NFW}}(x)=\ln(1+x)-x/(1+x)$, 
and $\Si(x)$ and $\Ci(x)$ are the sine and cosine integrals, respectively.
Eq.~(\ref{eq:lambdaknfw}) is normalized such that
the integral of $\lambda(x)$ over all space is unity.  
The systematic difference in halo profiles 
as a function of mass is encoded in the shift in 
halo concentrations at that mass.  
The power spectrum on scales of order 
$k \sim 1 \mpch$ is dominated by the one-halo term contribution from cluster-mass halos 
($\sim 10^{14} \hmsun$).  As such, we model the modification 
to the power spectrum due to cooling 
on scales near a few $\mpch$ by multiplying the dissipationless power spectrum by a ratio of 
Fourier-transformed NFW profiles, 
\begin{equation}
\label{eq:Pmodel}
P_{\mathrm{SF}}(k) \approx P_{\mathrm{DM}}(k)\ \Bigg[\frac{\lambda(\Rvir k/c_2,c_2)}{\lambda(\Rvir k/c_1,c_1)}\Bigg]^2.
\end{equation}
Roughly speaking, the second factor on the right side of 
Eq.~(\ref{eq:Pmodel}) represents a shift 
in the power spectrum due to a boost in the concentrations of halos which dominate the
largest nonlinear scales.  This factor is unity on scales larger 
than the typical scales of halos $k \lesssim \Rvir^{-1}$ 
and differs from unity at higher wavenumbers as one would expect for a one-halo effect.  
To implement this model, one chooses a characteristic $\Rvir$, a concentration $c_1$ 
that represents the halos in a typical dark matter simulation, and 
a second concentration $c_2$ that represents the effective concentrations of 
the mass profiles of halos with baryonic cooling and galaxy formation included.

While halo properties evolve, in our model of the
convergence power spectrum the lensing weight function has a broad
peak centered near $z \approx 0.55$, due to our choice of a single
thin sheet of sources at $z_s = 1$.  We therefore select parameters for
our model to represent typical halos at this
redshift, neglecting possible evolution.  At $z=0.55$, the largest
halos in our simulations have masses $\sim 10^{14} \hmsun$,
corresponding to virial radii of $\Rvir \sim 1.1 \hmpc$~comoving.  In the
dissipationless simulation, $\nbody$, halos of this mass have mean
concentrations $c_1 \sim 5$.  The concentrations in the $\csf$
simulation are enhanced by a factor of $\approx 1.7$ and so we take
$c_2 = 1.7 c_1 = 8.5$.

Figure~\ref{fig:model} shows the three-dimensional and convergence power
spectra that result from applying our model with these parameters
to the power spectra obtained from the $\nbody$ simulation. 
Notice several aspects of Figure~\ref{fig:model} which
have changed relative to Figure~\ref{fig:dpk} and Figure~\ref{fig:dcl}.  
First, we have removed the 
large-scale offset between the power spectra of the $\csf$ and $\nbody$ simulations
by forcing them to be equivalent at the lowest wavenumber, $k \approx 0.13 \mpch$.  
Second, in the right panel of Figure~\ref{fig:model}, we now plot 
convergence power spectra relative 
to the spectrum of the $\csf$ simulation rather than relative to the
$\nbody$ spectrum as in Fig.~\ref{fig:dcl}.  The $\nbody$ power spectrum now runs 
downward toward negative residual instead of the residual of the 
cooling run trending upward with wavenumber.  
For comparison, the dotted line in the left panel of Figure~\ref{fig:model} shows the result of 
adding a Poisson term to the dissipationless power spectrum as an alternative model 
for the net effect of cooling.  The motivation for a Poisson model is to 
treat the compact halo cores as point masses.  
Note that the Poisson model does not reproduce the correct power spectrum shape, 
regardless of amplitude.  Moreover, the large number densities required to produce 
even reasonable agreement ($\sim 0.1\ h^{3}$~Mpc$^{-3}$) in this regime correspond 
to halos of mass $\sim 10^{11} \hmsun$ which is significantly smaller 
than the halos that dominate the power on the range of scales we consider, and
thus the original physical motivation for the Poisson model is not substantiated.

The model of enhanced NFW concentrations, on the other hand,
successfully tracks the departure of the $\csf$ power
spectrum at scales $\ell \lesssim 4000$ from the $\nbody$ spectrum.
Thus we extend the range of $\ell$ where we can accurately model the power
spectrum by an order of magnitude at a cost of three extra parameters.
The model fails on small scales $\ell \gtrsim
10^4$ for at least two reasons.  The first is that the distributions of cold
gas and stellar mass are not well described by the high-concentration
NFW profile prescription within $\approx 0.05-0.1 \Rvir$, so the
model should break down when the central regions of halos contribute
significantly.  The second is that our choice of parameters
for this modification is driven by the properties of the largest halos
in our simulations.  This choice of parameters should not continue to
produce a viable model on scales where the structure of smaller halos
or of subhalos contributes significantly to the power.

Our fiducial choice of parameters $R_{\mathrm{vir}}$, $c_1$, and $c_2$
reflects the properties of the largest halos in the simulation
volume at redshift $z=0.55$.  In practice one should probably
introduce these parameters or a subset thereof (for example, one could
suppose that the concentrations of halos in dissipationless
simulations are well understood, fixing $c_1$) in any data analysis
and marginalize over our uncertainty in these nuisance parameters.
Ideally one would account for the evolution of these
parameters with redshift, perhaps by parameterizing this evolution or
by introducing independent sets of parameters for different tomographic
redshift bins.  It remains to be determined how much of the
constraining power from high-$\ell$ modes will be lost in this
marginalization process.  We explicate the evolution of these
parameters and the degradation of cosmological parameters arising from
the additional nuisance parameters in a forthcoming study.

Our results suggest that modifications to the power spectrum are
dominated by the one-halo term, which motivates this simple model.  We
have not attempted a more quantitative analysis because it is unclear
that such an analysis is warranted given the intrinsic limitations of
our simulation data.  Clearly, further theoretical efforts are needed
to constrain and understand the effects of galaxy formation physics
and refine the analytic model of these effects.  A first possible
refinement would be to scale dissipationless spectra by the ratio
of two complete one-halo terms [the form of which is given in
Eq.~(\ref{eq:P1h})].  In the upper term, the halos could follow a
different concentration-mass relation than dissipationless halos (for
example, parameterized by a different power law) in order to model the
modified structures of halos with baryonic effects included.  Another
possible refinement (at the cost of additional nuisance parameters)
would be to add additional components to the power spectrum modeling
to account for the distributions of hot and cold gas and stars within
halos and subhalos or to allow for more general profile shapes than NFW.

\section{Discussion and Conclusions}
\label{s:discussion}

We have performed a theoretical study of the impact of baryonic
physics on predictions of the matter power spectrum as it will be
measured by contemporary and forthcoming weak lensing surveys.  With the 
exception of the recent study by \citet[][]{jing_etal06}, such predictions have
hitherto been derived from dissipationless $N$-body simulations that
neglect the additional physics of the baryonic component of the
universe, or by semi-analytic models
\citep[{\it e.g.,}~][]{zhan_knox04,white04} which do not treat baryons
self-consistently.  Yet it is of paramount importance that these
predictions be extremely precise \citep[$\lesssim 1\%$,
e.g.][]{huterer_takada05} in order to realize fully the power of weak
lensing surveys to constrain the properties of dark energy.

We studied these effects using a suite of numerical simulations of
cosmological structure formation.  Each simulation started from the
same initial density field but differed in the treatment of 
baryons in the universe.  We also used the analytic and phenomenological
halo model of clustering in order to interpret our results and provide 
guidance for future studies.  Our primary results can be summarized as follows.

The effect of baryons on the matter power spectrum is significant at a
level that may be important for the interpretation of contemporary
weak lensing surveys and will almost certainly need to be accounted for 
in the interpretation of future surveys.  The enhancement in power in the
$\csf$ simulation relative to the $\nbody$ simulation due to baryon
dissipation and star formation at $\ell \gtrsim 800$ is large enough that it 
may affect the interpretation of the results from contemporary surveys like
the CFHT Wide Survey.  This enhancement is much larger than the
statistical error of future LSST- and SNAP-like surveys on scales that
have been used in previous forecasts of the constraining power of such
surveys ($\ell \lesssim 3 \times 10^3$).  We recognize that the stellar masses of 
the galaxies in our $\csf$ simulation are large compared to observed galaxies and 
take this to indicate that the enhancement that 
we quote for the $\csf$ simulation is likely to be an 
overestimate.  Nevertheless, we expect these effects to be quite important 
for future surveys.  Even in the absence of
cooling, future weak lensing surveys would still be sensitive to the
effects of different dynamics of dark matter and baryons and their
different final distributions in virialized halos.

Our findings are in qualitative agreement with the results of
\citet{jing_etal06}. However, there are quantitative differences,
which we attribute to different numerical hydrodynamics methods, and
different prescriptions for star formation and stellar feedback.  These 
differences serve to emphasize the fact that power spectrum predictions are sensitive
to the poorly-understood physics of galaxy formation and 
the specific details of the treatments of the baryonic components in 
different cosmological simulations.

The effects of baryons on the power spectrum on large scales ($k
\lesssim 1 \mpch$ or $\ell \lesssim 800$) are minimal.  In fact, we
find that {\em most of the identified effects on $P(k)$ are explained
by changes in the mass distributions within virialized objects in response
to baryonic physics.}  In the simulation which includes radiative
cooling, large amounts of gas cool and condense at halo centers where
they form stars.  As the baryons cool to halo centers, they drag dark
matter with them.  The net result is a large enhancement in the
relative concentration of matter toward the centers of halos resulting
in a dramatic increase in small-scale power.  On the other hand, the
gas in the $\ad$ simulations is unable to dissipate energy gained
during gravitational collapse.  Dark matter halos in this case have
their concentrations slightly enhanced leading to an increase in power
on intermediate scales ($1 \lesssim k/\mpch \lesssim 20$ or 
$800 \lesssim \ell \lesssim 10^4$), while the baryons associated with halos
remain hot and extended leading to a net deficit in power on small
scales ($k/\mpch \gtrsim 20$ or $\ell \gtrsim 10^4$).  We have used
detailed halo models of the matter power spectra designed to mimic our
simulation results in order to support this interpretation.

Observational measurements of the cluster concentration-mass relation
would be of significant help in constraining implementations of galaxy
formation physics in simulations and in testing the overall
adequacy of dissipationless simulations.  Such observational
constraints can be derived from either weak or strong lensing, or
through X-ray measurements of the intra-cluster medium \citep[see for
example,][]{buote_etal06,comerford_natarajan07}.  At present, these
measurements have their own sets of assumptions and systematic
uncertainties and their interpretation requires further tests against
numerical simulations. However, we envision that such
constraints can become significantly tighter in the future.

One of the primary points of this paper is that it is imperative that 
a discussion begin in earnest in order to determine how best to deal with 
the important, yet poorly-understood physics of galaxy formation in 
the context of forthcoming weak lensing surveys.  
In \S~\ref{sub:hmodel}, we showed how the halo model of clustering could be used to model the 
basic features of the matter power spectrum in the presence of baryons 
quite successfully and at a relatively minor cost.  
The model relies on a parameterized form 
for the one-halo term in $P(k)$, the parameters of which would be fit 
to either simulation or observational data as it becomes available.  
We stress that the modifications proposed in \S~\ref{sub:hmodel} work well 
to describe our simulations data, 
but we have not yet tested this approach at the requisite level of detail 
and on a wide enough spectrum of simulation results to have 
confidence that it will describe observational data precisely.  
Further, we have not studied to what degree the additional parameters 
required by this effective, parametric model of the effect of baryons 
on the matter power spectrum will degrade future constraints on 
dark energy parameters.  We plan to attack both of these issues in a 
forthcoming study.

As for the prospects for direct predictions of the matter power
spectrum at $k \gtrsim 1 \mpch$ or $\ell \gtrsim 10^3$, the challenge
facing theorists is and will remain significant.  Simulations which
treat the effects of baryons and galaxy formation physics are
currently limited in dynamic range.  As a consequence, simulations
that resolve small scales and follow processes relevant for galaxy
formation are unable to treat volumes that are large enough to render
cosmic variance insignificant and even these state-of-the-art simulations 
make assumptions about the net effects of processes that occur on 
scales that are unresolved.  Unlike dissipationless $N$-body
simulations, the computational cost of simulations that include these
additional physical processes is large enough that a direct
exploration of the way in which baryonic effects change with cosmology will be
impractical for some time.

Additionally, it is unclear whether current simulations with galaxy
formation have reached numerical convergence at the present achievable
resolution and with the included physics.  For example, current
simulations produce stellar fractions in galaxy cluster halos that are
a factor of two to three too large compared with observational
estimates of $\approx 15\%-20\%$ of the universal baryon fraction
\citep{lin_etal03,gonzalez_etal07}. Some of the
tension may be alleviated if a significant fraction of stars
are in a diffuse component not normally accounted for in observations
\citep[e.g., see][]{gonzalez_etal05,gonzalez_etal07,seigar_etal07}. There is 
additional, indirect evidence from observations of the intracluster
gas via X-rays and Sunyaev-Zel'dovich effects that the amount of
baryons condensing out of the hot gas is approximately correct in
simulations \citep{afshordi_etal07,nagai_etal07}.  Still, the role of
feedback from Active Galactic Nuclei or additional plasma physics in
halo cores and their effects on the distribution of baryons and the power
spectrum are also yet to be understood.

In summary, ab initio predictions for the matter power spectrum as a function 
of cosmology on relevant scales are not feasible, so alternative strategies 
must be explored.  We suggest that a worthwhile, general approach is to attempt
to isolate key effects of galaxy formation and encapsulate them in as
few parameters as possible. Such parameterized analytic models could be
used in conjunction with predictions of large-volume, dissipationless
simulations to model the observable convergence power spectra at the
expense of a small set of nuisance parameters. As we have shown in this work,
this approach appears to be quite promising if the primary effect of
galaxy formation processes relevant to convergence power spectra 
is to modify the distribution of mass within halos.

\vfill

\acknowledgements 

We would like to thank Hu Zhan for useful discussions on the possible
effects of baryon dissipation on the power spectrum during the early
stage of this study and for helpful comments on an early version of 
this manuscript.  We are grateful to Wayne Hu, Dragan Huterer, 
Jeremy Tinker, and Martin White for many useful discussions throughout the 
course of this project, to Sarah Hansen for several careful readings
of the manuscript, and to Anatoly Klypin for generating and providing initial conditions
and snapshots of the dissipationless simulation.  This project
was supported by the National Science Foundation (NSF) under grants
No. AST-0239759 and AST-0507666, by NASA through grant NAG5-13274, and
by the Kavli Institute for Cosmological Physics (KICP) at the
University of Chicago.  ARZ is supported by The NSF through the
Astronomy and Astrophysics Postdoctoral Fellowship Program under grant
AST-0602122 and by The KICP at The University of Chicago.  The
cosmological simulations used in this study were performed on the SGI
Altix system ({\tt columbia}) at NASA Ames and the IBM RS/6000 SP4
system ({\tt copper}) at the National Center for Supercomputing
Applications (NCSA).  In addition, we have made extensive use of the
NASA Astrophysics Data System and {\tt arXiv.org} preprint server.

\end{document}